\documentclass[aps,pra,twocolumn,superscriptaddress]{revtex4-2}
\usepackage{epsfig,amssymb, amsmath,fancyhdr,upgreek,color}
\newcommand{\bzeta}{{\boldsymbol \zeta}}
\newcommand{\bxi}{{\boldsymbol \xi}}
\newcommand{\bhata}{{\hat{\boldsymbol a}}}

\begin{document}
\title{End-to-End Fidelity and Rate Analysis for Zero-Added-Loss Multiplexing}
\author{Jeffrey H. Shapiro}
\email{jhs@mit.edu}
\affiliation{Department of Electrical Engineering and Computer Science, Research Laboratory of Electronics, Massachusetts Institute of Technology, Cambridge, Massachusetts 02139 USA}
\date{October 11, 2026}

\begin{abstract}
Zero-added-loss multiplexing (ZALM)~[Phys. Rev. Appl. {\bf 19}, 054029 (2023)] is a midpoint-source scheme for entanglement distribution. Its quantum transmitter (QTX) employs two Sagnac sources of polarization-entangled pairs plus partial Bell-state measurements (BSMs) on their idler outputs to herald entanglement in their signal outputs, thus avoiding the switching losses suffered by other multiplexing approaches.  When its Sagnac sources' spontaneous parametric downconverters (SPDCs) utilize $\chi^{(2)}$ crystals with phase-matched spectral islands, ZALM avoids the tradeoff between heralding efficiency and output-state fidelity incurred when conventional-crystal SPDCs' outputs are channelized by dense wavelength-division multiplexing filters.  This paper reports an end-to-end fidelity and rate analysis for islands-based ZALM that includes:  SPDC multipair emissions of all orders and partial-BSM losses in the ZALM QTX; loss in propagation from the ZALM QTX to Alice and Bob's quantum receivers (QRXs); loss and mode mismatch in the mode converters employed at Alice and Bob's QRXs; and loss in the partial-BSM loading of Alice and Bob's emissive quantum memories.   \end{abstract}

\maketitle

\section{Introduction \label{Intro}}
Sharing entanglement across long distances is essential to deployment of quantum network progenitors for a quantum internet~\cite{Lloyd2004,Kimble2008,Wehner2018}.  Although long-distance entanglement distribution has been demonstrated, i.e., 248\,km over deployed optical fiber~\cite{Neumann2022} and 1200\,km via a low-earth-orbit satellite~\cite{Yin2017}, these demonstrations' respective 9\,s$^{-1}$ and 1.1\,s$^{-1}$ distribution rates are many orders of magnitude below what is needed for even a rudimentary quantum network.  

Zero-added-loss multiplexing (ZALM), proposed by Chen~\emph{et al}.~\cite{Chen2023}, has emerged as a promising approach for dramatically increasing the long-distance delivery rate of high-fidelity entanglement.  ZALM's quantum transmitter (QTX) employs two Sagnac sources~\cite{Kim2006} of polarization-entangled pairs plus partial Bell-state measurements (BSMs)~\cite{Braunstein1995} on their idler outputs to herald entanglement in their signal outputs, thus avoiding the switching losses suffered by other multiplexing approaches~\cite{Mower2011,Dhara2022}.  When its Sagnac sources' spontaneous parametric downconverters (SPDCs) utilize $\chi^{(2)}$ crystals with phase-matched spectral islands~\cite{Morrison2022}, ZALM avoids the tradeoff between heralding efficiency and output-state fidelity incurred when conventional-crystal SPDCs' outputs are channelized by dense wavelength-division multiplexing (DWDM) filters~\cite{Shapiro2024}, as experienced in a recent experimental demonstration of ZALM with DWDM channelization~\cite{Tsujimoto2026}.  

To date, theoretical treatment of islands-based ZALM~\cite{Shapiro2025,Shapiro2026}  has included multipair emissions of all orders from the Sagnac sources' SPDCs, loss and dark counts in the QTX's partial BSM,  propagation loss from the QTX to Alice and Bob's quantum receivers (QRXs), and background light collected by those QRXs.  The present paper completes end-to-end fidelity and rate analysis for islands-based ZALM by extending prior work to include loss and mode mismatch in the mode converters needed to connect the arriving light to emissive color-center quantum memories~\cite{Knall2022,Bersin2024} at Alice and Bob's QRXs, as well as loss in the partial BSMs used to load those quantum memories.  For brevity's sake it ignores dark counts at both the QTX and QRXs, as well as background-light collection at the QRXs. These effects can be treated as done in Ref.~\cite{Shapiro2026}.  There it was shown that the low dark-count rates of available superconducting nanowire single-photon detectors lead to negligible dark-count impact on QTX performance.  Likewise, restricting the QRXs' light collection to a low time-bandwidth-product mode set results in negligible background-light impact on the Bell-state fidelity. 

The remainder of the paper is organized as follows.  Section~\ref{Setup} lays out the setup for islands-based ZALM, as analyzed in Refs.~\cite{Shapiro2025,Shapiro2026}, and augments that setup with models for Alice and Bob's emissive memories and their loading via partial BSMs.  Section~\ref{Primitives} provides the groundwork for obtaining the loaded-memories' density matrix by deriving transform-domain expressions for:  (1) the ZALM signal light arriving at Alice and Bob's QRXs following a QTX herald; (2) the joint state of the two memories' spins and emitted photons; and (3) the joint state that heralds the loading of Alice and Bob's memories.  Section~\ref{DensityMatrix} uses the preceding three primitives to derive the joint density matrix for Alice and Bob's loaded memories, a result used in Sec.~\ref{FidelityRate} to determine ZALM's end-to-end fidelity and entanglement-loading rate for transmission between its QTX and Alice and Bob's quantum memories.  Section~\ref{Discussion} wraps up the paper with some concluding discussion and suggestions for future work. 

\section{Islands-Based ZALM with Emissive Memories \label{Setup}}
Chen~\emph{et al}.'s ZALM proposal~\cite{Chen2023} presumed a satellite-borne QTX with free-space optics directing its heralded polarization entanglement to ground-based QRXs.  Chen~\emph{et al}.\@ did note, however, that a terrestrial ZALM system could transmit time-bin entanglement via optical fiber, see Chapman~\emph{et al}.~\cite{Chapman2026} for recent progress on a time-bin entanglement source for ZALM.  We too will assume that ZALM is distributing polarization entanglement but, unlike Ref.~\cite{Chen2023}, we will follow the lead of Shapiro~\emph{et al}.~\cite{Shapiro2025,Shapiro2026} by assuming the QTX's Sagnac SPDCs employ phase-matched spectral islands.  References~\cite{Shapiro2025,Shapiro2026} only treated the light arriving at Alice and Bob's QRXs following a QTX herald.  In this section we introduce models for Alice and Bob's emissive color-center quantum memories and their loading via partial BSMs, thus providing the setup for our end-to-end fidelity and rate analysis of islands-based ZALM.

The essentials of islands-based ZALM's QTX, as considered in Refs.~\cite{Shapiro2025,Shapiro2026}, are as follows.  The QTX's core is a pair of pulse-pumped, collinear Sagnac sources~\cite{footnote0}, $S_1$ and $S_2$, whose SPDCs each have $N_I$ identical, nonoverlapping, phase-matched spectral islands.  Each island emits spectrally factorable, i.e., single temporal mode, signal-idler photon pairs with center frequencies $\{\omega_{S_n},\omega_{I_n}: n = 1,2,\ldots, N_I\}$ in response to a single pump pulse.  Each pump pulse is assumed to provide equally-strong excitation to all $N_I$ spectral islands.  The joint signal-idler output state resulting from one such pump pulse is then the following tensor product of two-mode squeezed-vacuum states~\cite{footnote1}
\begin{align}
|\psi\rangle_{S_kI_k} = &\bigotimes_{n=1}^{N_I}\sum_{m=0}^\infty\sqrt{P_m}\,|m\rangle_{S_{k_{n_H}}}|m\rangle_{I_{k_{n_H}}} \nonumber\\[.05in]
&\bigotimes_{n'= 1}^{N_I}\sum_{m'=0}^\infty\sqrt{P_{m\prime}}\,|m'\rangle_{S_{k_{n'_V}}}|m'\rangle_{I_{k_{n'_V}}},
\label{islands_state}
\end{align}
for $k = 1,2$. 
Here: $|m\rangle_{K_{n_P}}$ is the $m$-photon Fock state of the $P$-polarized ($P=H,V$) signal ($K=S$) and idler ($K=I$) modes emitted by the $n$th phase-matched island; and 
\begin{equation}
P_m \equiv \frac{(G-1)^m}{G^{m+1}},\mbox{ for $m=0,1,\ldots, \infty$},
\label{BoseEinstein}
\end{equation} 
is the Bose-Einstein probability distribution with $G$ being the squeezing gain, so that $G-1 > 0$ is the average number of signal-idler pairs emitted per SPDC-island per pump pulse.  The joint signal-idler state of the two Sagnacs is then $|\psi\rangle_{{\bf SI}} = |\psi\rangle_{S_1I_1}|\psi\rangle_{S_2I_2}.$ 

\begin{figure}[hbt]
    \centering
\includegraphics[width=0.99\columnwidth]{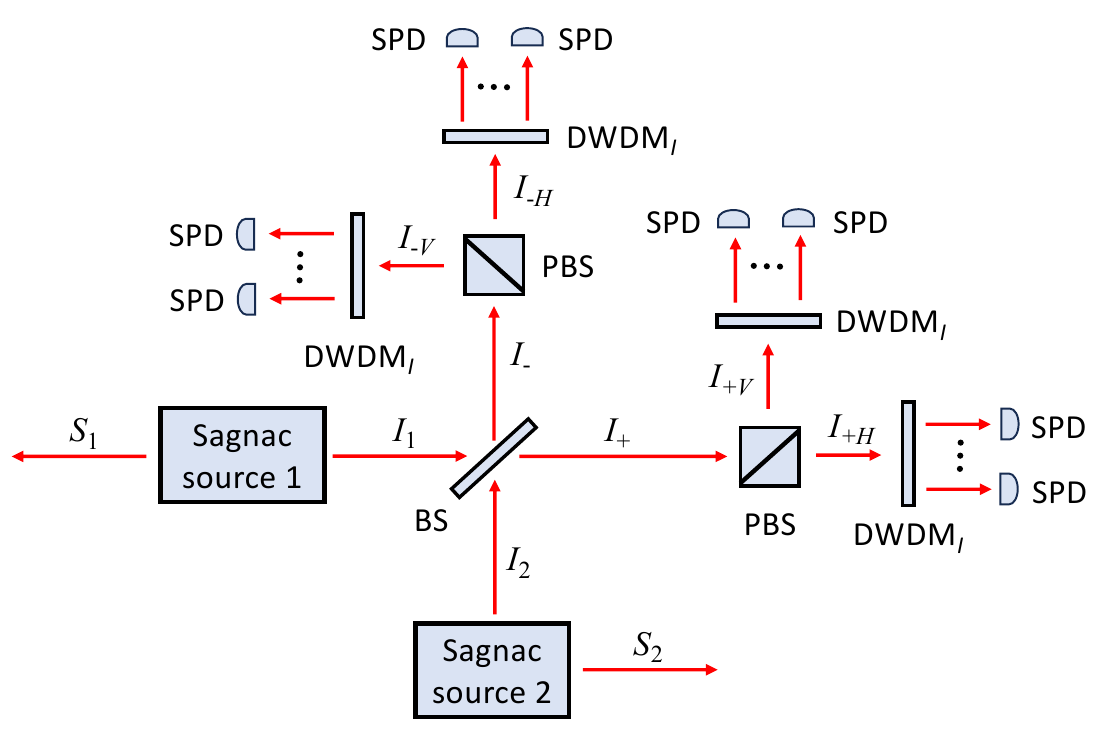}
    \caption{Schematic of islands-based ZALM's partial Bell-state measurement for heralding polarization-entangled photon pairs. Here $S_k$ and $I_k$ for $k = 1,2$ denote the signal ($S$) and idler ($I$) beams from the $k$th Sagnac source.  $I_\pm$ denote the idler-beam outputs from the 50--50 beam splitter (BS). $I_{\pm P}$ for $P= H, V$ denote the horizontally ($H$) and vertically ($V$) polarized outputs from the polarizing beam splitter (PBS) illuminated by $I_\pm$. DWDM$_I$:  idler-beam dense wavelength-division multiplexing filter.  SPD:  single-photon detector.
 \label{partial-BSM_fig} }
\end{figure}

The Sagnacs' idler-beam outputs, $I_1$ and $I_2$, are applied to the partial-BSM setup shown in Fig.~\ref{partial-BSM_fig}.  There, they: (1) are combined on a 50--50 beam splitter (BS) to erase which-path information; (2) are polarization analyzed by polarizing beam splitters (PBSs); and (3) have their spectral-island components separated by dense wavelength-division multiplexing (DWDM) filters.  Without appreciable loss of generality all losses in the QTX's idler and signal paths are assumed to be symmetrically distributed, with the former combined into an efficiency factor, $\eta_T$, ascribed to the partial-BSM's single-photon detectors (SPDs), and the latter lumped into the propagation transmissivity, $\eta_R$, to Alice and Bob's QRXs.  The SPDs will be taken to have $0, 1, >$1 partial photon-number resolution (PNR) capability and no dark counts.  A same-plus-cross-island (SPCI) herald from islands $n$ and $m$ occurs when single $H$-polarized and $V$-polarized photons are detected from the $n$th and $m$th islands, respectively.  When one detection occurs on the $I_+$ output from the 50--50 BS and the other from its $I_-$ output, a $\psi^-$ herald is declared.  Alternatively,  when both occur from the $I_+$ or the $I_-$ output, a $\psi^+$ herald is declared.  The $S_1$ and $S_2$ signal outputs from the two Sagnacs are sent to Alice and Bob along with a classical transmission containing the herald information, i.e., $\psi^-$ or $\psi^+$, and the island indices $n$ and $m$.   Alice and Bob use this heralding information to attempt loading the heralded Bell state,
\begin{equation}
|\psi^\mp\rangle_{\tilde{\bf S}}  = \frac{|H\rangle_{\tilde{S}_{A_n}}|V\rangle_{\tilde{S}_{B_m}}\mp |V\rangle_{\tilde{S}_{A_m}}|H\rangle_{{\tilde{S}_{B_n}}}}{\sqrt{2}},
\end{equation} 
where $\tilde{\bf S} = \tilde{S}_A\tilde{S}_B$ are the signal modes arriving at Alice and Bob's QRXs.  
Note that we say ``a herald is declared'' because half of the heralds are \emph{false} heralds, in which photons are sent to Alice or Bob but \emph{not} to both, hence only one QRX's quantum memory can be loaded if no background light is collected.

It follows from Eq.~(\ref{islands_state}) that all $nm$th-islands $\psi^-$ heralds lead to the same QTX output state and likewise for all $nm$th-islands $\psi^+$ heralds.  Thus we shall drop island indices in what follows.  Moreover, until Sec.~\ref{DensityMatrix} we shall limit our attention to the $\psi^-$ herald from $I'_{+H}I'_{-V}$ single-photon detections, where the $I'$ symbols indicate that the QTX efficiency $\eta_T$ is accounted for.  Shapiro~\emph{et al}.~\cite{Shapiro2025}, using a transform-domain approach, derived the conditional density operator for the  $\tilde{{\bf S}}$ modes arriving at Alice and Bob's QRXs after propagation from the ZALM QTX through transmissivity-$\eta_R$ channels, obtaining
\begin{equation}
\hat{\rho}_{\tilde{{\bf S}}\mid I'_{+H}I'_{-V}} = \int\!\frac{{\rm d}^8\bzeta_{\tilde{S}}}{\pi^4}\,\chi_A^{\rho_{\tilde{{\bf S}}\mid I'_{+H}I'_{-V}}}(\bzeta_{\tilde{S}})e^{-\bhata_{\tilde{S}}^\dagger\bzeta_{\tilde{S}}}e^{\bzeta_{\tilde{S}}^\dagger\bhata_{\tilde{S}}},
\label{preprimitive1}
\end{equation}
where
\begin{align}
&\chi_A^{\rho_{\tilde{{\bf S}}\mid I'_{+H}I'_{-V}}}(\bzeta_{\tilde{S}}) = e^{-\bzeta_{\tilde{S}}^\dagger\bzeta_{\tilde{S}}/N_S'}\nonumber \\[.05in]
&\times \left[1-\frac{\eta_R|\zeta_{\tilde{S}_{A_H}}+\zeta_{\tilde{S}_{B_H}}|^2}{2N_S}\right]\!\left[1-\frac{\eta_R|\zeta_{\tilde{S}_{A_V}}-\zeta_{\tilde{S}_{B_V}}|^2}{2N_S}\right],
\label{primitive1}
\end{align}
with $N_S\equiv [\eta_T(G-1)+1]/G$ and $N_S'\equiv [\eta_R/N_S+(1-\eta_R)]^{-1}$, 
is the density operator's anti-normally ordered characteristic function.  In these equations
\begin{equation}
\bzeta^\dagger_{\tilde{S}} \equiv \left[\begin{array}{cccc}\zeta^*_{\tilde{S}_{A_H}} & \zeta^*_{\tilde{S}_{A_V}} & \zeta^*_{\tilde{S}_{B_H}} & \zeta^*_{\tilde{S}_{B_V}}\end{array}\right], 
\end{equation}
and
\begin{equation}
\bhata_{\tilde{S}}^\dagger \equiv \left[\begin{array}{cccc}\hat{a}^\dagger_{\tilde{S}_{A_H}} & \hat{a}^\dagger_{\tilde{S}_{A_V}} & \hat{a}^\dagger_{\tilde{S}_{B_H}} & \hat{a}^\dagger_{\tilde{S}_{B_V}} \end{array}\right],
\end{equation}
are four-dimensional (4D) row vectors of complex-valued parameters and photon creation operators, respectively, and the integral is from $-$$\infty$ to $\infty$ in all its dimensions.  Also, as was the case in Eq.~(\ref{islands_state}), the density operator and characteristic function from Eqs.~(\ref{preprimitive1}) and (\ref{primitive1}) only apply to the signals' single temporal-mode excitations for the islands in question, while the other temporal modes are in their vacuum states.  So, because we will be allowing for mode mismatch between the arriving light and the memories' emitted light in our fidelity and rate analysis, it behooves us at this point to introduce the relevant temporal modes in conjunction with our models for the QRXs' mode converters and emissive color-center quantum memories.

The ZALM QTX will likely transmit $\sim$10\,ps duration, 1.5\,$\upmu$m wavelength signal photons to Alice and Bob's QRXs, whose emissive quantum memories will likely use Si-vacancy color centers that emit $\sim$ns duration, 737\,nm wavelength photons~\cite{Chen2023}.  The QRXs' mode converters must therefore do quantum-state wavelength conversion, bandwidth compression, and mode shaping~\cite{footnote2} of the arriving light to efficiently couple the arriving light to the quantum memories.  All signal-beam losses, i.e., in the ZALM QTX, in propagation to Alice and Bob's QRXs, and in the mode conversion process, can (and will) be lumped together into $\eta_R$.  Hence, without loss of generality, we take Eqs.~(\ref{preprimitive1}) and (\ref{primitive1}) to apply to the signals' single-temporal mode excitations at the output of Alice and Bob's mode converters, viz., the inputs to their quantum memories' partial BSMs.  To do so only requires that the excited modes used in conjunction with those equations are those at the memories' inputs.  Let us see how these excited modes relate to those emitted by the quantum memories.

Suppose that $\Phi_S(t)e^{-i\omega_{S_n}t}$ for horizontal polarization, and $\Phi_S(t)e^{-i\omega_{S_m}t}$ for vertical polarization, are the normalized ($\int\!{\rm d}t\,|\Phi_S(t)|^2 = 1$) positive-frequency excited modes arriving at Alice and Bob's QRXs from an $nm$th-island herald.  Now suppose that Alice and Bob's quantum memories emit photons in the $\phi_E(t)e^{-i\omega_Et}$ normalized ($\int\!{\rm d}t\,|\phi_E(t)|^2 = 1$) positive-frequency  mode.  After the arriving light undergoes polarization-dependent wavelength conversion,  bandwidth compression, and mode shaping, suppose that it is in the $\phi_S(t)e^{-i\omega_Et}$ normalized ($\int\!{\rm d}t\,|\phi_S(t)|^2 = 1$) positive-frequency mode. A Gram-Schmidt procedure enables us quantify mode mismatch with the addition of only one auxiliary mode.  In particular, with
\begin{equation}
\gamma \equiv \int\!{\rm d}t\,\phi_S^*(t)\phi_E(t)
\label{gamma}
\end{equation}
and 
\begin{equation}
\phi^\perp(t) \equiv \frac{\phi_E(t) - \gamma\phi_S(t)}{\sqrt{1-|\gamma|^2}},
\end{equation}
we have that
\begin{equation}
\phi_E(t) = \gamma\phi_S(t) + \sqrt{1-|\gamma|^2}\,\phi_S^\perp(t),
\end{equation}
with $\phi_S(t)$ and $\phi_S^\perp(t)$ being orthonormal.  Note that mode mismatch ($|\gamma|^2 <1$) can be caused by imperfect wavelength conversion, or by insufficient bandwidth compression, or by shape mismatch, or by any combination thereof.  

To complete the foundation for what follows, it only remains for us to provide models for Alice and Bob's quantum memories and the partial-BSMs they use to load them.  In keeping with our assumption of polarization, rather than time-bin, entanglement distribution, we shall assume that these quantum memories produce the joint spin-photon pure state
\begin{equation}
|\psi\rangle_{M_KE_K} = \frac{|\uparrow\rangle_{M_K}|H\rangle_{E_K} + |\downarrow\rangle_{M_K}|V\rangle_{E_K}}{\sqrt{2}},
\end{equation}
for $K = A,B$, where the photonic states are for the $\phi_E(t)e^{-i\omega_Et}$ positive-frequency modes with all other temporal modes being in their vacuum states.  All losses in coupling the emitted photons to their respective quantum memories will be accounted for by a sub-unit success probability, $p_c$, for each memory.  

\begin{figure}[hbt]
    \centering
\includegraphics[width=0.95\columnwidth]{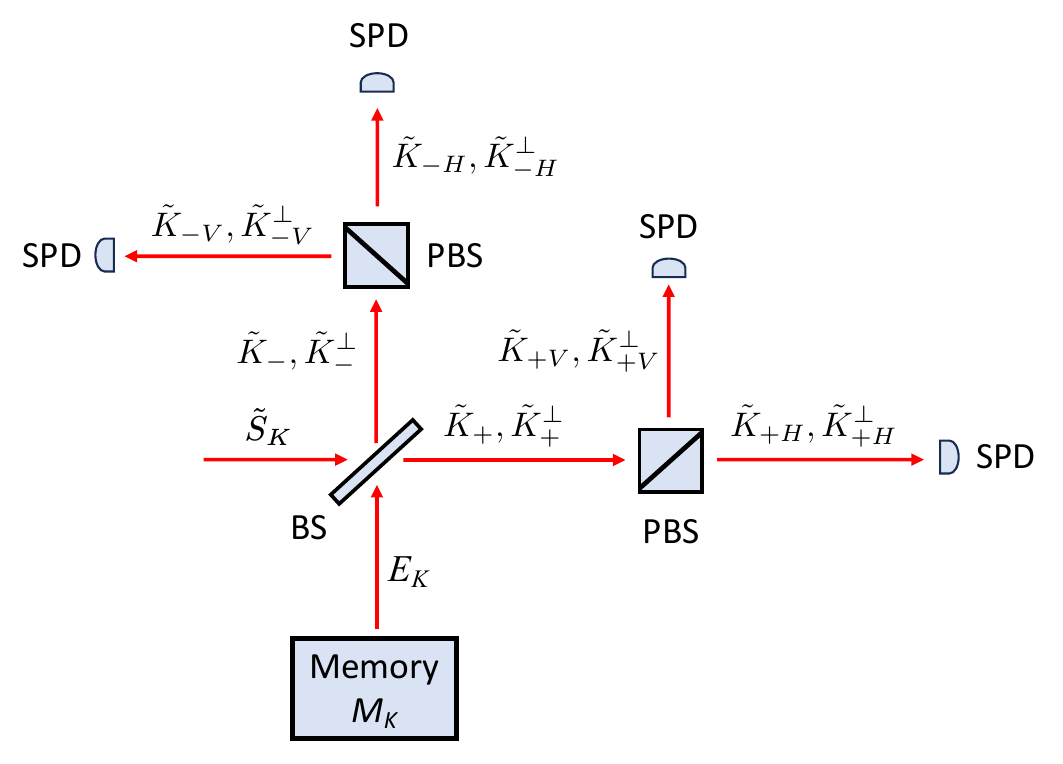}
    \caption{Schematic of partial Bell-state measurement for heralding quantum memory loading at Alice ($K = A$) and Bob's ($K=B$) QRXs.  Only excited or potentially-excited modes are shown. $\tilde{S}_K$:  post mode-conversion arriving signal light, its positive-frequency mode shape is $\phi_S(t)e^{-i\omega_Et}$. $E_K$:  light emitted by quantum memory, its positive-frequency mode shape is $\phi_E(t)e^{-i\omega_Et}$.  $\tilde{E}_\pm$ and $\tilde{E}^\perp_\pm$:  outputs from 50--50 beam splitter (BS), with their positive-frequency mode shapes being $\phi_S(t)e^{-i\omega_Et}$ and $\phi_S^\perp(t)e^{-i\omega_Et}$, respectively. $\tilde{E}_{\pm P}$ and $\tilde{E}^\perp_{\pm P}$ for $P = H,V$: polarization-resolved outputs from transmitting $\tilde{E}_\pm$ and $\tilde{E}^\perp_\pm$ through polarizing beam splitters (PBSs).  SPD: single-photon detector.
 \label{memory-BSM_fig} }
\end{figure}

Figure~\ref{memory-BSM_fig} shows a schematic for the partial BSMs used by Alice ($K=A$) and Bob ($K=B$) to herald loading of their quantum memories.  They function in the same manner as the ZALM QTX's partial-BSM from Fig.~\ref{partial-BSM_fig}, i.e., loading of memory $K$ is heralded when the partial-PNR (0, 1, $>$1) SPDs record single-photon detections in one $H$-polarized and one $V$-polarized channel.  Losses in these BSM setups will be lumped together into sub-unit quantum efficiencies, $\eta_D$, for their SPDs, just as we did with $\eta_T$ for the ZALM QTX's partial BSM. Unlike the QTX's partial BSM, half of whose heralds are false heralds even in the absence of dark counts, there are \emph{no} false heralds at the quantum memories in the absence of background light and dark counts.  This is so because with probability $p_c$, which will appear in the rate analysis but not in our fidelity calculation, there will be a memory-emitted photon entering Fig.~\ref{memory-BSM_fig}'s 50--50 beam splitter.   Also unlike the ZALM QTX's partial BSM, whose SPDs are only illuminated by single-mode photons in the absence of background light, Alice and Bob's partial BSMs will be illuminated by two potentially-excited modes, even in the absence of background light, when there is mode mismatch ($|\gamma|^2 <1$).  As a result we need to be careful in defining what constitutes single $H$-polarized and $V$-polarized detections, as we now show.   

Assuming that the duration of the SPDs' detection interval includes the entire durations of their illumination's excited modes, then a single-photon detection at a particular SPD implies a detection from either its $\phi_S(t)e^{-i\omega_Et}$ or its $\phi_S^\perp(t)e^{-i\omega_Et}$ mode, but not both.  For example, a set of single-photon detections that heralds loading of both Alice and Bob's memories is
\begin{displaymath}
(\tilde{A}'_{+H}\oplus\tilde{A}^{\perp\prime}_{+H}) \cup
(\tilde{A}'_{-V}\oplus\tilde{A}^{\perp\prime}_{-V}) \cup
(\tilde{B}'_{+H}\oplus\tilde{B}^{\perp\prime}_{+H}) \cup
(\tilde{B}'_{-V}\oplus\tilde{B}^{\perp\prime}_{-V}),
\end{displaymath}
where the $\tilde{A}',\tilde{B}'$ symbols indicate that the sub-unit partial-BSM efficiency, $\eta_D$, has been applied, and $\oplus$ denotes exclusive or.  To avoid this cumbersome notation in what follows, we will use the shorthand $\mathcal{A}'_{+H}\mathcal{A}'_{-V}\mathcal{B}'_{+H}\mathcal{B}'_{-V}$ in its place. \\ 

\section{Primitives for Fidelity and Rate Analysis \label{Primitives}}
This section's task is to derive transform-domain expressions that will serve as primitives from which the loaded-memories density matrix will be obtained.  Before proceeding, it is worth noting the potential complexity to be faced.  There are four possible photon-detection patterns that generate a partial-BSM herald, and end-to-end ZALM operation with emissive memories employs three partial BSMs.  Consequently, there are $64 = 4^3$ source-memory heralds whose resulting loaded-memories' density matrix we must ultimately obtain to complete our fidelity and rate analysis.  In this section we will only address one of them:  an $I'_{+H}I'_{-V}$ QTX herald and an $\mathcal{A}'_{+H}\mathcal{A}'_{-V}\mathcal{B}'_{+H}\mathcal{B}'_{-V}$ emissive memories herald.   

\subsection{Arriving-Light Primitive}
The first primitive we seek is the anti-normally ordered characteristic function of the relevant signal modes entering Alice and Bob's partial-BSM setups.  Equation~(\ref{primitive1}) already gives us the characteristic function for the $\tilde{\bf S} = \tilde{S}_A\tilde{S}_B$ modes, but we adjoin to it the characteristic function for the $\tilde{\bf S}^\perp = \tilde{S}_A^\perp\tilde{S}_B^\perp$ modes because of possible mode mismatch.  The result is
\begin{equation}
\chi_A^{\rho_{\tilde{\bf S}\mid I'_{+H}I'_{-V}}}(\bzeta_{\tilde{S}})\,{\rm Tr}\!\left[|{\bf 0}\rangle_{\tilde{S}^\perp}{}_{\tilde{S}^\perp}\langle {\bf 0}|e^{-\bzeta_{\tilde{S}^\perp}^\dagger\bhata_{\tilde{S}^\perp}}e^{\bhata_{\tilde{S}^\perp}^\dagger\bzeta_{\tilde{S}^\perp}}\right]
\label{primitive1'}
\end{equation}
where $\chi_A^{\rho_{\tilde{\bf S}\mid I'_{+H}I'_{-V}}}(\bzeta_{\tilde{S}})$ is given by Eq.~(\ref{primitive1}) and 
\begin{equation}
{\rm Tr}\!\left[|{\bf 0}\rangle_{\tilde{S}^\perp}{}_{\tilde{S}^\perp}\langle {\bf 0}|e^{-\bzeta_{\tilde{S}^\perp}^\dagger\bhata_{\tilde{S}^\perp}}e^{\bhata_{\tilde{S}^\perp}^\dagger\bzeta_{\tilde{S}^\perp}}\right] = e^{-\bzeta_{\tilde{S}^\perp}^\dagger \bzeta_{\tilde{S}^\perp}}.
\end{equation}

As will be seen in Sec.~\ref{DensityMatrix},  a more useful form of the arriving-light  primitive is obtained by recognizing that the memories' partial-BSM efficiency commutes with Fig.~\ref{memory-BSM_fig}'s polarizing beam splitters and 50--50 beam splitter.  Applying $\eta_D$ to the $\tilde{S}_K$ and $\tilde{S}_K^\perp$ modes yields, respectively, the $\tilde{S}'_K$ and $\tilde{S}^{\perp \prime}_K$ modes whose characteristic function is our first primitive,
\begin{widetext}
\begin{align}
\mbox{Primitive 1 } = e^{-\bzeta_{\tilde{S}'}^\dagger\bzeta_{\tilde{S}'}/N_D}e^{-\bzeta_{\tilde{S}^{\perp\prime}}^\dagger\bzeta_{\tilde{S}^{\perp \prime}}}\!\left[1-\frac{\eta_D\eta_R|\zeta_{\tilde{S}'_{AH}}+\zeta_{\tilde{S}'_{BH}}|^2}{2N_S}\right]\!\left[1-\frac{\eta_D\eta_R|\zeta_{\tilde{S}'_{AV}}-\zeta_{\tilde{S}'_{BV}}|^2}{2N_S}\right],
\end{align}
\end{widetext}  
where $N_D \equiv [\eta_D/N_S' + (1-\eta_D)]^{-1}$.

\subsection{Emissive-Memories' Primitive}
The second primitive we require is the photonic modes' anti-normally ordered characteristic function for Alice and Bob's memory spins and emitted photons.  This partial transform can be constructed from their individual partial transforms, viz., ${\rm Tr}_{E_K}\!\left[|\psi\rangle_{M_KE_K}{}_{M_KE_K}\langle\psi| e^{-\bzeta_ {E_K}^\dagger\bhata_{E_K}}e^{\bhata_{E_K}^\dagger\bzeta_{E_K}}\right], $ for $K = A,B$. Using the Baker-Campbell-Hausdorff formula to normally order $e^{-\bzeta_ {E_K}^\dagger\bhata_{E_K}}e^{\bhata_{E_K}^\dagger\bzeta_{E_K}}$~\cite{Nielsen2000}, we get
\begin{align}
&{\rm Tr}_{E_K}\!\left[|\psi\rangle_{M_KE_K}{}_{M_KE_K}\langle\psi| e^{-\bzeta_ {E_K}^\dagger\bhata_{E_K}}e^{\bhata_{E_K}^\dagger\bzeta_{E_K}}\right] =
\nonumber \\[.05in]
&\,\, {\rm Tr}_{E_K}\!\left[|\psi\rangle_{M_KE_K}{}_{M_KE_K}\langle\psi| e^{-\bzeta_{E_K}^\dagger\bzeta_{E_K}} e^{\bhata_{E_K}^\dagger\bzeta_{E_K}}e^{-\bzeta_ {E_K}^\dagger\bhata_{E_K}}\right]. 
\end{align}
Using the cyclic nature of the trace, and the first two Taylor-series terms for $e^{\bhata_{E_K}^\dagger\bzeta_{E_K}}$ and $e^{-\bzeta_ {E_K}^\dagger\bhata_{E_K}}$, these matrices are easily found to be 
\begin{align}
&\frac{e^{-\bzeta_{E_K}^\dagger\bzeta_{E_K}}}{2} \nonumber \\[.05in]
&\,\times \left[\begin{array}{cc} (|\uparrow\rangle_{M_K}{}_{M_K}\langle \uparrow|)T_{\uparrow K\uparrow K} & (|\uparrow\rangle_{M_K}{}_{M_K}\langle \downarrow|)T_{\uparrow K \downarrow K}\\[.05in] 
(|\downarrow\rangle_{M_K}{}_{M_K}\langle \uparrow|)T_{\downarrow K 
\uparrow K}& (|\downarrow\rangle_{M_K}{}_{M_K}\langle \downarrow|)T_{\downarrow K\downarrow K}\end{array} \right],
\end{align}
where
\begin{align}
T_{\uparrow K \uparrow K} & = 1-|\zeta_{E_{KH}}|^2\\[.05in]
T_{\uparrow K \downarrow K} & = -\zeta_{E_{KH}}\zeta^*_{E_{KV}}\\[.05in]
T_{\downarrow K \uparrow K} & = -\zeta^*_{E_{KH}}\zeta_{E_{KV}}\\[.05in]
T_{\downarrow K \downarrow K}& = 1-|\zeta_{E_{KV}}|^2.  
\end{align}
Of course, because the preceding matrices apply to the $\phi_E(t)e^{-i\omega_Et}$ modes, we must rewrite them in terms of the $\phi_S(t) e^{-i\omega_Et}$ and $\phi_S^\perp(t)e^{-i\omega_Et}$ modes to deal with mode mismatch.  The result so obtained is 
\begin{align}
&\frac{e^{-\bzeta_{\tilde{E}_K}^\dagger\bzeta_{\tilde{E}_K}}e^{-\bzeta_{\tilde{E}^\perp_K}^\dagger\bzeta_{\tilde{E}^\perp_K}}}{2}\nonumber \\[.05in] &\times\left[\begin{array}{cc} (|\uparrow\rangle_{M_K}{}_{M_K}\langle \uparrow|)\tilde{T}_{\uparrow K\uparrow K} & (|\uparrow\rangle_{M_K}{}_{M_K}\langle \downarrow|)\tilde{T}_{\uparrow K \downarrow K}\\[.05in] 
(|\downarrow\rangle_{M_K}{}_{M_K}\langle \uparrow|)\tilde{T}_{\downarrow K 
\uparrow K}& (|\downarrow\rangle_{M_K}{}_{M_K}\langle \downarrow|)\tilde{T}_{\downarrow K\downarrow K}\end{array} \right], \nonumber
\end{align}
with
\begin{align}
\tilde{T}_{\uparrow_K\uparrow_K} &= 1-|\gamma\zeta^*_{\tilde{E}_{KH}} + \sqrt{1-|\gamma|^2}\,\zeta^*_{\tilde{E}^\perp_{KH}}|^2\\[.05in]
\tilde{T}_{\uparrow_K\downarrow_K} &= -(\gamma^*\zeta_{\tilde{E}_{KH}} + \sqrt{1-|\gamma|^2}\zeta_{\tilde{E}^\perp_{KH}}) \nonumber \\[.05in]
&\, \times (\gamma\zeta^*_{\tilde{E}_{KV}}+\sqrt{1-|\gamma|^2}\,\zeta^*_{\tilde{E}^\perp_{KV}})\\[.05in]
\tilde{T}_{\downarrow_K\uparrow_K} &= -(\gamma^*\zeta_{\tilde{E}_{KV}} + \sqrt{1-|\gamma|^2}\,\zeta_{\tilde{E}^\perp_{KV}})\nonumber \\[.05in]
&\,\times (\gamma\zeta^*_{\tilde{E}_{KH}}+\sqrt{1-|\gamma|^2}\,\zeta^*_{\tilde{E}^\perp_{KH}})\\[.05in]
\tilde{T}_{\downarrow_K\downarrow_K} &= 1-|\gamma\zeta^*_{\tilde{E}_{KV}} + \sqrt{1-|\gamma|^2}\,\zeta^*_{\tilde{E}^\perp_{KV}}|^2.  
\end{align}

As was done for the arriving-light primitive, we will commute the memories' partial-BSM efficiency with the polarizing and 50--50 beam splitters for reasons that will become apparent in the next section.  So, applying $\eta_D$ to transform the $\tilde{E}_K$ and $\tilde{E}^\perp_K$ modes to the $\tilde{E}'_K$ and $\tilde{E}^{\perp \prime}_K$ modes we get
\begin{align}
&\frac{e^{-\bzeta_{\tilde{E}'_K}^\dagger\bzeta_{\tilde{E}'_K}}e^{-\bzeta_{\tilde{E}^{\perp\prime}_K}^\dagger\bzeta_{\tilde{E}^{\perp\prime}_K}}}{2}\nonumber \\[.05in] &\times\left[\begin{array}{cc} (|\uparrow\rangle_{M_K}{}_{M_K}\langle \uparrow|)\tilde{T}'_{\uparrow K\uparrow K} & (|\uparrow\rangle_{M_K}{}_{M_K}\langle \downarrow|)\tilde{T}'_{\uparrow K \downarrow K}\\[.05in] 
(|\downarrow\rangle_{M_K}{}_{M_K}\langle \uparrow|)\tilde{T}'_{\downarrow K 
\uparrow K}& (|\downarrow\rangle_{M_K}{}_{M_K}\langle \downarrow|)\tilde{T}'_{\downarrow K\downarrow K}\end{array} \right], \nonumber
\end{align}
where
\begin{align}
\tilde{T}'_{\uparrow_K\uparrow_K} &= 1-\eta_D|\gamma\zeta^*_{\tilde{E}'_{KH}} + \sqrt{1-|\gamma|^2}\,\zeta^*_{\tilde{E}^{\perp\prime}_{KH}}|^2 \label{state_coeff11}\\[.05in]
\tilde{T}'_{\uparrow_K\downarrow_K} &= -\eta_D(\gamma^*\zeta_{\tilde{E}'_{KH}} + \sqrt{1-|\gamma|^2}\zeta_{\tilde{E}^{\perp\prime}_{KH}}) \nonumber \\[.05in]
&\, \times (\gamma\zeta^*_{\tilde{E}'_{KV}}+\sqrt{1-|\gamma|^2}\,\zeta^*_{\tilde{E}^{\perp\prime}_{KV}}) \label{state_coeff12}\\[.05in]
\tilde{T}'_{\downarrow_K\uparrow_K} &= -\eta_D(\gamma^*\zeta_{\tilde{E}'_{KV}} + \sqrt{1-|\gamma|^2}\,\zeta_{\tilde{E}^{\perp\prime}_{KV}})\nonumber \\[.05in]
&\,\times (\gamma\zeta^*_{\tilde{E}'_{KH}}+\sqrt{1-|\gamma|^2}\,\zeta^*_{\tilde{E}^{\perp\prime}_{KH}}) \label{state_coeff21}\\[.05in]
\tilde{T}'_{\downarrow_K\downarrow_K} &= 1-\eta_D|\gamma\zeta^*_{\tilde{E}'_{KV}} + \sqrt{1-|\gamma|^2}\,\zeta^*_{\tilde{E}^{\perp\prime}_{KV}}|^2. \label{state_coeff22}  
\end{align}

Finally, we combine the preceding results for $K = A,B$ and get the core of the memories' unnormalized density matrix, i.e., Primitive~2, given by
\begin{widetext}
\begin{equation}
\mbox{Primitive~2 } = \frac{e^{-\bzeta_{\tilde{E}}^\dagger\bzeta_{\tilde{E}}}e^{-\bzeta_{\tilde{E}^\perp}^\dagger\bzeta_{\tilde{E}^\perp}}}{4} \left[\begin{array}{cccc}
\tilde{T}'_{\uparrow_A\uparrow_A}\tilde{T}'_{\uparrow_B\uparrow_B}& \tilde{T}'_{\uparrow_A\uparrow_A}\tilde{T}'_{\uparrow_B\downarrow_B}& \tilde{T}'_{\uparrow_A\downarrow_A}\tilde{T}'_{\uparrow_B\uparrow_B}&\tilde{T}'_{\uparrow_A\downarrow_A}\tilde{T}'_{\uparrow_B\downarrow_B}\\[.05in]
\tilde{T}'_{\uparrow_A\uparrow_A}\tilde{T}'_{\downarrow_B\uparrow_B} & \tilde{T}'_{\uparrow_A\uparrow_A}\tilde{T}'_{\downarrow_B\downarrow_B}& \tilde{T}'_{\uparrow_A\downarrow_A}\tilde{T}'_{\downarrow_B\uparrow_B} & \tilde{T}'_{\uparrow_A\downarrow_A}\tilde{T}'_{\downarrow_B\downarrow_B}\\[.05in]
\tilde{T}'_{\downarrow_A\uparrow_A}\tilde{T}'_{\uparrow_B\uparrow_B} & \tilde{T}'_{\downarrow_A\uparrow_A}\tilde{T}'_{\uparrow_B\downarrow_B}& \tilde{T}'_{\downarrow_A\downarrow_A}\tilde{T}'_{\uparrow_B\uparrow_B} & \tilde{T}'_{\downarrow_A\downarrow_A}\tilde{T}'_{\uparrow_B\downarrow_B}\\[.05in]
\tilde{T}'_{\downarrow_A\uparrow_A}\tilde{T}'_{\downarrow_B\uparrow_B} & \tilde{T}'_{\downarrow_A\uparrow_A}\tilde{T}'_{\downarrow_B\downarrow_B} & \tilde{T}'_{\downarrow_A\downarrow_A}\tilde{T}'_{\downarrow_B\uparrow_B}  & \tilde{T}'_{\downarrow_A\downarrow_A}\tilde{T}'_{\downarrow_B\downarrow_B}
\end{array}\right].
\end{equation}
\end{widetext}

\subsection{Memories'-Herald Primitive}
The product of Primitive~1 and Primitive~2 is the photonic modes' anti-normally ordered characteristic function of the arriving light, the photons emitted by Alice and Bob's quantum memories, and those memories' spins, under the assumptions that there has been an $I'_{+H}I'_{-V}$ herald at the ZALM QTX, and the effect of the memories' sub-unit partial-BSM efficiency has been included. To get the spins' unnormalized density matrix under those assumptions when there has been an $\mathcal{A}'_{+H}\mathcal{A}'_{-V}\mathcal{B}'_{+H}\mathcal{B}'_{-V}$ memories herald we need a third primitive:  the transform-domain representation of that herald backpropagated to before  Fig.~\ref{memory-BSM_fig}'s 50--50 beam splitter. From 
\begin{align}
\Pr&(\tilde{A}'_{+H} = n) = {}_{\tilde{A}'_{+H}}\langle n|\hat{\rho}_{\tilde{A}'_{+H}}|n\rangle_{\tilde{A}'_{+H}} \\[.05in]
&= \int\!\frac{{\rm d}^2\zeta_{\tilde{A}'_{+H}}}{\pi}\,\chi_A^{\rho_{\tilde{A}'_{+H}}}(\zeta_{\tilde{A}'_{+H}}) \nonumber \\[.05in]
&\, \times{}_{\tilde{A}'_{+H}}\langle n| e^{-\zeta_{\tilde{A}'_{+H}}\hat{a}^{\prime \dagger}_{\tilde{A}'_{+H}}} e^{\zeta^*_{\tilde{A}'_{+H}}\hat{a}'_{\tilde{A}'_{+H}}}|n\rangle_{A'_{+H}}\\[.05in]
& = \int\!\frac{{\rm d}^2\zeta_{\tilde{A}'_{+H}}}{\pi}\,\chi_A^{\rho_{\tilde{A}'_{+H}}}(\zeta_{\tilde{A}'_{+H}})(1-|\zeta_{\tilde{A}'_{+H}}|^2)^n,
\end{align}
for $n= 0,1$, and similar results for other modes, it follows that 
\begin{align}
&[(1-|\zeta_{\tilde{A}'_{+H}}|^2)+(1-|\zeta_{\tilde{A}^{\perp\prime}_{+H}}|^2)] \nonumber \\[.05in]
&\,\times[(1-|\zeta_{\tilde{A}'_{-V}}|^2)+(1-|\zeta_{\tilde{A}^{\perp\prime}_{-V}}|^2)] \nonumber \\[.05in]
&\,\times
[(1-|\zeta_{\tilde{B}'_{+H}}|^2)+(1-|\zeta_{\tilde{B}^{\perp\prime}_{+H}}|^2)] \nonumber\\[.05in]
&\,\times[(1-|\zeta_{\tilde{B}'_{-V}}|^2)+(1-|\zeta_{\tilde{B}^{\perp\prime}_{-V}}|^2)]
\end{align}
is the transform-domain representation of the $\mathcal{A}'_{+H}\mathcal{A}'_{-V}\mathcal{B}'_{+H}\mathcal{B}'_{-V}$ memories herald at Fig.~\ref{memory-BSM_fig}'s SPDs.  Backpropagating this result across that figure's polarizing and 50--50 beam splitters we arrive at Primitive~3:
\begin{align}
&\mbox{Primitive~3 } = \nonumber \\[.05in]
&\left[\left(1-\frac{|\zeta_{\tilde{S}'_{AH}}+\zeta_{\tilde{E}'_{AH}}|^2}{2}\right) + 
\left(1-\frac{|\zeta_{\tilde{S}^{\perp\prime}_{AH}}+\zeta_{\tilde{E}^{\perp\prime}_{AH}}|^2}{2}\right)\right]\nonumber \\[.05in]
&\times \left[\left(1-\frac{|\zeta_{\tilde{S}'_{AV}}-\zeta_{\tilde{E}'_{AV}}|^2}{2}\right) + 
\left(1-\frac{|\zeta_{\tilde{S}^{\perp\prime}_{AV}}-\zeta_{\tilde{E}^{\perp\prime}_{AV}}|^2}{2}\right)\right] \nonumber
\end{align}
\begin{align}
&\times \left[\left(1-\frac{|\zeta_{\tilde{S}'_{BH}}+\zeta_{\tilde{E}'_{BH}}|^2}{2}\right) + 
\left(1-\frac{|\zeta_{\tilde{S}^{\perp\prime}_{BH}}+\zeta_{\tilde{E}^{\perp\prime}_{BH}}|^2}{2}\right)\right] \nonumber\\[.05in]
&\times \left[\left(1-\frac{|\zeta_{\tilde{S}'_{BV}}-\zeta_{\tilde{E}'_{BV}}|^2}{2}\right) + 
\left(1-\frac{|\zeta_{\tilde{S}^{\perp\prime}_{BV}}-\zeta_{\tilde{E}^{\perp\prime}_{BV}}|^2}{2}\right)\right].
\end{align}

For the assumed QTX and QRXs' heralding events,
\begin{align}
\tilde{\rho}_{\bf M} &= \int\!\frac{{\rm d}^8\bzeta_{\tilde{S}'}}{\pi^4}\int\!\frac{{\rm d}^8\bzeta_{\tilde{E}'}}{\pi ^4}\int\!\frac{{\rm d}^8\bzeta_{\tilde{S}^{\perp\prime}}}{\pi^4}\int\!\frac{{\rm d}^8\bzeta_{\tilde{E}^{\perp\prime}}}{\pi ^4} \nonumber \\[.05in]
& \,\,\times \left(\prod_{i=1}^3\mbox{Primitive $i$}\right)
\label{rhoM1} 
\end{align}
is the $4\times 4$ unnormalized density matrix for Alice and Bob's loaded memories, viz., the quantity we need for our fidelity and rate analysis.  Note that the expression on the right is a 32D-dimensional integral over real variables.  Furthermore, Eq.~(\ref{rhoM1})  only applies to a single source-memories herald.  There are 63 more source-memories heralds whose unnormalized density matrices we must evaluate to complete our fidelity and rate analysis.  Thankfully, the primitives we have in hand will make that task quite manageable.  

\section{Loaded-Memories' Density Matrix \label{DensityMatrix}}
For the $I'_{+H}I'_{-V}$ source herald and $\mathcal{A}'_{+H}\mathcal{A}'_{-V}\mathcal{B}'_{+H}\mathcal{B}'_{-V}$ memories herald, Eq.~(\ref{rhoM1})'s unnormalized density matrix for the loaded memories is
\begin{widetext}
\begin{equation}
\tilde{\rho}_{\bf M} = 
\left[\begin{array}{cccc}
\tilde{\rho}_{|\uparrow_A\uparrow_B\rangle \langle \uparrow_A\uparrow_B|} & \tilde{\rho}_{|\uparrow_A\uparrow_B\rangle \langle \uparrow_A\downarrow_B|}& \tilde{\rho}_{|\uparrow_A\uparrow_B\rangle \langle \downarrow_A\uparrow_B|} & \tilde{\rho}_{|\uparrow_A\uparrow_B\rangle \langle \downarrow_A\downarrow_B|} \\[.05in]
\tilde{\rho}_{|\uparrow_A\downarrow_B\rangle \langle \uparrow_A\uparrow_B|} & \tilde{\rho}_{|\uparrow_A\downarrow_B\rangle \langle \uparrow_A\downarrow_B|}& \tilde{\rho}_{|\uparrow_A\downarrow_B\rangle \langle \downarrow_A\uparrow_B|} & \tilde{\rho}_{|\uparrow_A\downarrow_B\rangle \langle \downarrow_A\downarrow_B|} \\[.05in]
\tilde{\rho}_{|\downarrow_A\uparrow_B\rangle \langle \uparrow_A\uparrow_B|} & \tilde{\rho}_{|\downarrow_A\uparrow_B\rangle \langle \uparrow_A\downarrow_B|}& \tilde{\rho}_{|\downarrow_A\uparrow_B\rangle \langle \downarrow_A\uparrow_B|} & \tilde{\rho}_{|\downarrow_A\uparrow_B\rangle \langle \downarrow_A\downarrow_B|} \\[.05in]
\tilde{\rho}_{|\downarrow_A\downarrow_B\rangle \langle \uparrow_A\uparrow_B|} & \tilde{\rho}_{|\downarrow_A\downarrow_B\rangle \langle \uparrow_A\downarrow_B|}& \tilde{\rho}_{|\downarrow_A\downarrow_B\rangle \langle \downarrow_A\uparrow_B|} & \tilde{\rho}_{|\downarrow_A\downarrow_B\rangle \langle \downarrow_A\downarrow_B|}
\end{array}\right]
\end{equation}
in the standard basis, which we will rewrite in a more compact shorthand as
\begin{equation}
\tilde{\rho}_{\bf M} = \left[\begin{array}{cccc}
\tilde{\rho}_{11} & \tilde{\rho}_{12} & \tilde{\rho}_{13} & \tilde{\rho}_{14} \\[.05in]
\tilde{\rho}_{21} & \tilde{\rho}_{22} & \tilde{\rho}_{23} & \tilde{\rho}_{24} \\[.05in]
\tilde{\rho}_{31} & \tilde{\rho}_{32} & \tilde{\rho}_{33} & \tilde{\rho}_{34} \\[.05in]
\tilde{\rho}_{41} & \tilde{\rho}_{42} & \tilde{\rho}_{43} & \tilde{\rho}_{44} \end{array}\right].
\end{equation}
Because $\tilde{\rho}_{\bf M}$ is Hermitian, only 10 of its elements need to be found.  All of them are of the form
\begin{align}
\int\!\frac{{\rm d}^8\bzeta_{\tilde{S}'}}{\pi^4}\int\!\frac{{\rm d}^8\bzeta_{\tilde{E}'}}{\pi ^4}
&\int\!\frac{{\rm d}^8\bzeta_{\tilde{S}^{\perp\prime}}}{\pi^4}\int\!\frac{{\rm d}^8\bzeta_{\tilde{E}^{\perp\prime}}}{\pi ^4}
\,e^{-\bzeta_{\tilde{S}'}^\dagger\bzeta_{\tilde{S}'}/N_D}e^{-\bzeta_{\tilde{E}'}^\dagger\bzeta_{E'}} 
e^{-\bzeta_{\tilde{S}^{\perp\prime}}^\dagger\bzeta_{\tilde{S}^{\perp\prime}}}e^{-\bzeta_{\tilde{E}^{\perp\prime}}^\dagger\bzeta_{\tilde{E}^{\perp\prime}}} \nonumber \\[.05in]
&\,\times\,
{\rm poly}_{ij}(\bzeta_{\tilde{S}'},\bzeta_{\tilde{E}'},\bzeta^*_{\tilde{S}'},\bzeta^*_{\tilde{E}'},\bzeta_{\tilde{S}^{\perp\prime}},\bzeta_{\tilde{E}^{\perp\prime}},\bzeta^*_{\tilde{S}^{\perp\prime}},\bzeta^*_{\tilde{E}^{\perp\prime}}),
\label{polyintegral}
\end{align}
\end{widetext}
where ${\rm poly}_{ij}(\bzeta_{\tilde{S}'},\bzeta_{\tilde{E}'},\bzeta^*_{\tilde{S}'},\bzeta^*_{\tilde{E}'},\bzeta_{\tilde{S}^{\perp\prime}},\bzeta_{\tilde{E}^{\perp\prime}},\bzeta^*_{\tilde{S}^{\perp\prime}},\bzeta^*_{\tilde{E}^{\perp\prime}})$ is a polynomial of its vector arguments' components.  Defining fictitious complex-valued Gaussian random vectors $\bxi_{\tilde{S}'}, \bxi_{\tilde{S}^\perp\prime},\bxi_{\tilde{E}'},\bxi_{\tilde{E}^\perp\prime}$ by the probability density functions (pdfs)
\begin{align}
p_{\bxi_{\tilde{S}'}}(\bzeta_{\tilde{S}'}) &\equiv \frac{e^{-\bzeta_{\tilde{S}'}^\dagger\bzeta_{\tilde{S}'}/N_D}}{(\pi N_D)^4},\,\,
p_{\bxi_{\tilde{E}'}}(\bzeta_{\tilde{E}'}) \equiv \frac{e^{-\bzeta_{\tilde{E}'}^\dagger\bzeta_{\tilde{E}'}}}{\pi^4}\nonumber \\[.05in]
p_{\bxi_{\tilde{S}^{\perp\prime}}}(\bzeta_{\tilde{S}^{\perp\prime}}) &\equiv \frac{e^{-\bzeta_{\tilde{S}^{\perp\prime}}^\dagger\bzeta_{\tilde{S}^{\perp\prime}}}}{\pi^4},\,\,
p_{\bxi_{\tilde{E}^{\perp\prime}}}(\bzeta_{\tilde{E}^{\perp\prime}}) \equiv \frac{e^{-\bzeta_{\tilde{E}^{\perp\prime}}^\dagger\bzeta_{\tilde{E}^{\perp\prime}}}}{\pi^4},
\end{align}
we see that the $\tilde{\rho}_{ij}$ can be evaluated by Gaussian moment factoring.  Indeed, because these random vectors are statistically independent, circulo-complex Gaussian random vectors, that moment factoring will be easily accomplished, as we now show.

Let us begin with the $\tilde{\rho}_{\bf M}$'s diagonal elements. Using $\mathbb{E}(\cdot)$ to denote expectation over the fictitious random vectors $\bxi_{\tilde{S}'},\bxi_{\tilde{S}^{\perp\prime}},\bxi_{\tilde{E}'}$ and $\bxi_{\tilde{E}^{\perp\prime}}$ we have that
\begin{widetext}
\begin{align}
&\tilde{\rho}_{11} = N_D^4\,\mathbb{E}\!\left\{ 
\left[1-\frac{\eta_D\eta_R(|\xi_{\tilde{S}'_{AH}}|^2+|\xi_{\tilde{S}'_{BH}}|^2)}{2N_S}\right]\!\left[1-\frac{\eta_D\eta_R(|\xi_{\tilde{S}'_{AV}}|^2+|\xi_{\tilde{S}'_{BV}}|^2)}{2N_S}\right] 
\right. \nonumber \\[.05in]
&\!\!\times [1-\eta_D(|\gamma|^2|\xi_{\tilde{E}'_{AH}}|^2 + (1-|\gamma|^2)|\xi_{\tilde{E}^{\perp\prime}_{AH}}|^2)][1-\eta_D(|\gamma|^2|\xi_{\tilde{E}'_{BH}}|^2 + (1-|\gamma|^2)|\xi_{\tilde{E}^{\perp\prime}_{BH}}|^2)]/4 \nonumber \\[.05in]
&\!\!\times 
\left[\!\left(1-\frac{|\xi_{\tilde{S}'_{AH}}|^2+|\xi_{\tilde{E}'_{AH}}|^2}{2}\right)+ \left(1-\frac{|\xi_{\tilde{S}^{\perp\prime}_{AH}}|^2 + |\xi_{\tilde{E}^{\perp\prime}_{AH}}|^2}{2}\right)\!\right]\!\left[\!\left(1-\frac{|\xi_{\tilde{S}'_{AV}}|^2+|\xi_{\tilde{E}'_{AV}}|^2}{2}\right) \!+\!\left(1-\frac{|\xi_{\tilde{S}^{\perp\prime}_{AV}}|^2+|\xi_{\tilde{E}^{\perp\prime}_{AV}}|^2}{2}\right)\!\right] \nonumber \\[.05in]
&\!\!\times \left. \left[\!\left(1-\frac{|\xi_{\tilde{S}'_{BH}}|^2+|\xi_{\tilde{E}'_{BH}}|^2}{2}\right)+ \left(1-\frac{|\xi_{\tilde{S}^{\perp\prime}_{BH}}|^2 + |\xi_{\tilde{E}^{\perp\prime}_{BH}}|^2}{2}\right)\!\right]\!\left[\!\left(1-\frac{|\xi_{\tilde{S}'_{BV}}|^2+|\xi_{\tilde{E}'_{BV}}|^2}{2}\right) \!+\! \left(1-\frac{|\xi_{\tilde{S}^{\perp\prime}_{BV}}|^2+|\xi_{\tilde{E}^{\perp\prime}_{BV}}|^2}{2}\right)\!\right]\!\right\},
\label{rhotilde11}
\end{align}
\begin{align}
&\tilde{\rho}_{22} = N_D^4\,\mathbb{E}\!\left\{ 
\left[1-\frac{\eta_D\eta_R(|\xi_{\tilde{S}'_{AH}}|^2+|\xi_{\tilde{S}'_{BH}}|^2)}{2N_S}\right]\!\left[1-\frac{\eta_D\eta_R(|\xi_{\tilde{S}'_{AV}}|^2+|\xi_{\tilde{S}'_{BV}}|^2)}{2N_S}\right] 
\right. \nonumber \\[.05in]
&\!\!\times [1-\eta_D(|\gamma|^2|\xi_{\tilde{E}'_{AH}}|^2 + (1-|\gamma|^2)|\xi_{\tilde{E}^{\perp\prime}_{AH}}|^2)][1-\eta_D(|\gamma|^2|\xi_{\tilde{E}'_{BV}}|^2 + (1-|\gamma|^2)|\xi_{\tilde{E}^{\perp\prime}_{BV}}|^2)]/4 \nonumber \\[.05in]
&\!\!\times 
\left[\!\left(1-\frac{|\xi_{\tilde{S}'_{AH}}|^2+|\xi_{\tilde{E}'_{AH}}|^2}{2}\right)+ \left(1-\frac{|\xi_{\tilde{S}^{\perp\prime}_{AH}}|^2 + |\xi_{\tilde{E}^{\perp\prime}_{AH}}|^2}{2}\right)\!\right]\!\left[\!\left(1-\frac{|\xi_{\tilde{S}'_{AV}}|^2+|\xi_{\tilde{E}'_{AV}}|^2}{2}\right) \!+\!\left(1-\frac{|\xi_{\tilde{S}^{\perp\prime}_{AV}}|^2+|\xi_{\tilde{E}^{\perp\prime}_{AV}}|^2}{2}\right)\!\right] \nonumber \\[.05in]
&\!\!\times \left. \left[\!\left(1-\frac{|\xi_{\tilde{S}'_{BH}}|^2+|\xi_{\tilde{E}'_{BH}}|^2}{2}\right)+ \left(1-\frac{|\xi_{\tilde{S}^{\perp\prime}_{BH}}|^2 + |\xi_{\tilde{E}^{\perp\prime}_{BH}}|^2}{2}\right)\!\right]\!\left[\!\left(1-\frac{|\xi_{\tilde{S}'_{BV}}|^2+|\xi_{\tilde{E}'_{BV}}|^2}{2}\right) \!+\! \left(1-\frac{|\xi_{\tilde{S}^{\perp\prime}_{BV}}|^2+|\xi_{\tilde{E}^{\perp\prime}_{BV}}|^2}{2}\right)\!\right]\!\right\},
\label{rhotilde22}
\end{align}
\begin{align}
&\tilde{\rho}_{33} = N_D^4\,\mathbb{E}\!\left\{ 
\left[1-\frac{\eta_D\eta_R(|\xi_{\tilde{S}'_{AH}}|^2+|\xi_{\tilde{S}'_{BH}}|^2)}{2N_S}\right]\!\left[1-\frac{\eta_D\eta_R(|\xi_{\tilde{S}'_{AV}}|^2+|\xi_{\tilde{S}'_{BV}}|^2)}{2N_S}\right] 
\right. \nonumber \\[.05in]
&\!\!\times [1-\eta_D(|\gamma|^2|\xi_{\tilde{E}'_{AV}}|^2 + (1-|\gamma|^2)|\xi_{\tilde{E}^{\perp\prime}_{AV}}|^2)][1-\eta_D(|\gamma|^2|\xi_{\tilde{E}'_{BH}}|^2 + (1-|\gamma|^2)|\xi_{\tilde{E}^{\perp\prime}_{BH}}|^2)]/4 \nonumber \\[.05in]
&\!\!\times 
\left[\!\left(1-\frac{|\xi_{\tilde{S}'_{AH}}|^2+|\xi_{\tilde{E}'_{AH}}|^2}{2}\right)+ \left(1-\frac{|\xi_{\tilde{S}^{\perp\prime}_{AH}}|^2 + |\xi_{\tilde{E}^{\perp\prime}_{AH}}|^2}{2}\right)\!\right]\!\left[\!\left(1-\frac{|\xi_{\tilde{S}'_{AV}}|^2+|\xi_{\tilde{E}'_{AV}}|^2}{2}\right) \!+\!\left(1-\frac{|\xi_{\tilde{S}^{\perp\prime}_{AV}}|^2+|\xi_{\tilde{E}^{\perp\prime}_{AV}}|^2}{2}\right)\!\right] \nonumber \\[.05in]
&\!\!\times \left. \left[\!\left(1-\frac{|\xi_{\tilde{S}'_{BH}}|^2+|\xi_{\tilde{E}'_{BH}}|^2}{2}\right)+ \left(1-\frac{|\xi_{\tilde{S}^{\perp\prime}_{BH}}|^2 + |\xi_{\tilde{E}^{\perp\prime}_{BH}}|^2}{2}\right)\!\right]\!\left[\!\left(1-\frac{|\xi_{\tilde{S}'_{BV}}|^2+|\xi_{\tilde{E}'_{BV}}|^2}{2}\right) \!+\! \left(1-\frac{|\xi_{\tilde{S}^{\perp\prime}_{BV}}|^2+|\xi_{\tilde{E}^{\perp\prime}_{BV}}|^2}{2}\right)\!\right]\!\right\},
\label{rhotilde33}
\end{align}
and
\begin{align}
&\tilde{\rho}_{44} = N_D^4\,\mathbb{E}\!\left\{ 
\left[1-\frac{\eta_D\eta_R(|\xi_{\tilde{S}'_{AH}}|^2+|\xi_{\tilde{S}'_{BH}}|^2)}{2N_S}\right]\!\left[1-\frac{\eta_D\eta_R(|\xi_{\tilde{S}'_{AV}}|^2+|\xi_{\tilde{S}'_{BV}}|^2)}{2N_S}\right] 
\right. \nonumber \\[.05in]
&\!\!\times [1-\eta_D(|\gamma|^2|\xi_{\tilde{E}'_{AV}}|^2 + (1-|\gamma|^2)|\xi_{\tilde{E}^{\perp\prime}_{AV}}|^2)][1-\eta_D(|\gamma|^2|\xi_{\tilde{E}'_{BV}}|^2 + (1-|\gamma|^2)|\xi_{\tilde{E}^{\perp\prime}_{BV}}|^2)]/4 \nonumber \\[.05in]
&\!\!\times 
\left[\!\left(1-\frac{|\xi_{\tilde{S}'_{AH}}|^2+|\xi_{\tilde{E}'_{AH}}|^2}{2}\right)+ \left(1-\frac{|\xi_{\tilde{S}^{\perp\prime}_{AH}}|^2 + |\xi_{\tilde{E}^{\perp\prime}_{AH}}|^2}{2}\right)\!\right]\!\left[\!\left(1-\frac{|\xi_{\tilde{S}'_{AV}}|^2+|\xi_{\tilde{E}'_{AV}}|^2}{2}\right) \!+\!\left(1-\frac{|\xi_{\tilde{S}^{\perp\prime}_{AV}}|^2+|\xi_{\tilde{E}^{\perp\prime}_{AV}}|^2}{2}\right)\!\right] \nonumber \\[.05in]
&\!\!\times \left. \left[\!\left(1-\frac{|\xi_{\tilde{S}'_{BH}}|^2+|\xi_{\tilde{E}'_{BH}}|^2}{2}\right)+ \left(1-\frac{|\xi_{\tilde{S}^{\perp\prime}_{BH}}|^2 + |\xi_{\tilde{E}^{\perp\prime}_{BH}}|^2}{2}\right)\!\right]\!\left[\!\left(1-\frac{|\xi_{\tilde{S}'_{BV}}|^2+|\xi_{\tilde{E}'_{BV}}|^2}{2}\right) \!+\! \left(1-\frac{|\xi_{\tilde{S}^{\perp\prime}_{BV}}|^2+|\xi_{\tilde{E}^{\perp\prime}_{BV}}|^2}{2}\right)\!\right]\!\right\},
\label{rhotilde44}
\end{align}
\end{widetext}
because all the cross terms---like those in $|\xi_{{\tilde{S}}'_{AH}}+\xi_{\tilde{E}'_{AH}}|^2$---vanish after averaging due to $\bxi_{\tilde{S}'},\bxi_{\tilde{S}^{\perp\prime}},\bxi_{\tilde{E}'}$, and $\bxi_{\tilde{E}^{\perp\prime}}$ being independent, circulo-complex, Gaussian random vectors.  To see that this is so, first multiply out all the terms in ${\rm poly}_{ii}$.  Then perform the integrations for obtaining $\tilde{\rho}_{ii}$ from Eq~(\ref{polyintegral}) in the following order:  first the $\bzeta_{\tilde{S}^{\perp\prime}}$ integral, second the $\bzeta_{\tilde{E}^{\perp\prime}}$ integral, third the $\bzeta_{\tilde{E}'}$ integral, and finally the $\bzeta_{\tilde{S}'}$ integral.  

At this point we note that $\tilde{\rho}_{11} = \tilde{\rho}_{44}$, because Eqs.~(\ref{rhotilde11}) and (\ref{rhotilde22}) differ by an exchange $H$ for $V$ in primitive~2 and the $H$ and $V$ components of $\bxi_{\tilde{\bf E}'}$ and $\bxi_{\tilde{\bf E}^{\perp\prime}}$ are identically distributed.  A similar argument shows that $\tilde{\rho}_{22} = \tilde{\rho}_{33}$.  Then, because the squared magnitude of a circulo-complex random variable is exponentially distributed, we can easily complete the Gaussian moment factoring that gives the diagonal elements we are seeking:
\begin{align}
&\tilde{\rho}_{11} = \tilde{\rho}_{44}  
=\{(N_D-1) N_D^4 [1 -(1-\eta_D) N_D]/64N_S^2\} \nonumber\\[.05in]
&\,\,\times [2 \eta_D \eta_R N_D^2 + N_S - 
    N_D (\eta_D \eta_R + N_S)]\nonumber \\[.05in]
&\,\,\times [-N_S +
     N_D (\eta_D \eta_R - 
       2 (1-\eta_D) \eta_D \eta_R N_D \nonumber \\[.05in]
       &\,\, + (1-\eta_D) N_S)],
       \label{diag11}
\end{align}
and
\begin{align}
\tilde{\rho}_{22} &= \tilde{\rho}_{33} = (N_D^4/256 N_S^2) \{\eta_D \eta_R N_D \nonumber \\[.05in]
&\,\,\times[-2 + 
N_D (6 - 3 \eta_D - 
4 (1 - \eta_D) N_D))] \nonumber \\[.05in]
&\,\, + 2 (1 - N_D) (1 - (1 - \eta_D) N_D) N_S\}^2.
\label{diag22}
\end{align}

Turning now to the density matrix's off-diagonal terms, we do Eq.~(\ref{polyintegral})'s integrals in the same order we used for the $\tilde{\rho}_{ii}$ and find that
\begin{equation}
\tilde{\rho}_{12} = \tilde{\rho}_{13} = \tilde{\rho}_{14} = \tilde{\rho}_{24} = \tilde{\rho}_{34} = 0,
\end{equation}
and
\begin{equation}
\tilde{\rho}_{23} = \tilde{\rho}_{32} = -|\gamma|^4\eta_D^4\eta_R^2N_D^8/256N_S^2.
\label{offdiag}
\end{equation}

Thus, for the $I'_{+H}I'_{-V}$ source herald and the $\mathcal{A}'_{+H}\mathcal{A}'_{-V}\mathcal{B}'_{+H}\mathcal{B}'_{-V}$ memories herald we have that the loaded-memories' unnormalized density matrix is
\begin{equation}
\tilde{\rho}_{\bf M} = \left[\begin{array}{cccc}
\tilde{\rho}_{11} & 0 & 0 & 0 \\[.05in]
0 & \tilde{\rho}_{22} & \tilde{\rho}_{23} & 0 \\[.05in]
0 & \tilde{\rho}_{23} & \tilde{\rho}_{22} & 0 \\[.05in]
0 & 0 & 0 & \tilde{\rho}_{11} \end{array}\right],
\label{densmatstructure}
\end{equation}
where we have used its Hermitian nature and $\tilde{\rho}_{23}$'s being real valued.  Several points are worth making at this time.  First, the density matrix's diagonal elements do not depend on the mode mismatch, i.e., they are independent of $\gamma$.  Second, it is diagonal in the Bell basis, i.e., $\{|\psi^-\rangle_{\bf  M},|\psi^+\rangle_{\bf M},|\phi^+\rangle_{\bf M},|\phi^-\rangle_{\bf M}\}$, where
\begin{equation}
|\psi^\mp\rangle_{\bf M}  = \frac{|\uparrow\rangle_{M_A}|\downarrow\rangle_{M_B}\mp |\downarrow\rangle_{M_A}|\uparrow\rangle_{M_B}}{\sqrt{2}},
\end{equation} 
and
\begin{equation}
|\phi^\pm\rangle_{\bf M}  = \frac{|\uparrow\rangle_{M_A}|\uparrow\rangle_{M_B}\pm |\downarrow\rangle_{M_A}|\downarrow\rangle_{M_B}}{\sqrt{2}}.
\end{equation} 
Specifically, the unitary that transforms the standard basis to the Bell basis,
\begin{equation}
U \equiv \frac{1}{\sqrt{2}}\!\left[\begin{array}{cccc} 
0 & 1 & -1 & 0\\[.05in]
0 & 1 & 1 & 0\\[.05in]
1& 0 & 0 & 1\\[.05in]
1 & 0 & 0 &-1\end{array}\right],
\end{equation}
gives us
\begin{equation}
\tilde{\rho}^{(B)}_{\bf M} = U\tilde{\rho}_{\bf M}U^\dagger  = 
\!\left[\begin{array}{cccc}
\tilde{\rho}_{22}-\tilde{\rho}_{23} & 0 & 0 & 0\\[.05in]
0 & \tilde{\rho}_{22}+\tilde{\rho}_{23} & 0 & 0\\[.05in]
0 & 0 & \tilde{\rho}_{11} & 0 \\[.05in]
0 & 0 & 0 & \tilde{\rho}_{11}\end{array}\right]
\label{rhoB}
\end{equation}
for the unnormalized density matrix in the Bell basis.  Furthermore, $\tilde{\rho}_{23}$'s being negative for $|\gamma|>0$ implies that the Bell-basis density matrix's $\psi^-$ element exceeds its $\psi^+$ element.  Interestingly, it turns out that $\tilde{\rho}_{22} + \tilde{\rho}_{23} = \tilde{\rho}_{11}$ in the absence of mode mismatch, implying that the normalized density matrix, $\hat{\rho}_{\bf M}$, is a Werner state when $|\gamma|^2 = 1$.

To complete our preparation for fidelity and rate analysis we need $\tilde{\rho}_{\bf M}$ for the remaining 63 source-memories heralds.  Finding them is easier than one might think.  Sticking with the $I'_{+H}I'_{-V}$ source herald, the following 8 memories heralds,
\begin{equation}
\begin{array}{c}
\mathcal{A}'_{+H}\mathcal{A}'_{-V}\mathcal{B}'_{+H}\mathcal{B}'_{-V}, \,\mathcal{A}'_{+H}\mathcal{A}'_{-V}\mathcal{B}'_{-H}\mathcal{B}'_{+V}, \\[.05in] 
\mathcal{A}'_{-H}\mathcal{A}'_{+V}\mathcal{B}'_{+H}\mathcal{B}'_{-V}, \,\mathcal{A}'_{-H}\mathcal{A}'_{+V}\mathcal{B}'_{-H}\mathcal{B}'_{+V}, \\[.05in]
\mathcal{A}'_{+H}\mathcal{A}'_{+V}\mathcal{B}'_{+H}\mathcal{B}'_{+V}, \,\mathcal{A}'_{+H}\mathcal{A}'_{+V}\mathcal{B}'_{-H}\mathcal{B}'_{-V}, \\[.05in] 
\mathcal{A}'_{-H}\mathcal{A}'_{-V}\mathcal{B}'_{+H}\mathcal{B}'_{+V}, \,\mathcal{A}'_{-H}\mathcal{A}'_{-V}\mathcal{B}'_{-H}\mathcal{B}'_{-V} \end{array} 
\nonumber
\end{equation} 
only effects on the ${\rm poly}_{ij}$ is to flip some signs in primitive~3's cross terms.  It follows that their density matrices all have the same diagonal elements.  Moreover, because the parities of their cross-term signs are all even, their density operators have the same off-diagonal elements.  Thus, the results given earlier in this section now apply to 8 source-memories heralds.  For these same 8 memories heralds, changing the source herald to $I'_{+V}I'_{-H}$ produces even-parity phase flips in primitive~1's cross terms.  Consequently, the resulting density matrices are the same as what we have already found.  

Now consider the 8 remaining memories' heralds, viz.,
\begin{align}
\begin{array}{c}
\mathcal{A}'_{+H}\mathcal{A}'_{+V}\mathcal{B}'_{+H}\mathcal{B}_{-V},\, \mathcal{A}'_{+H}\mathcal{A}'_{+V}\mathcal{B}'_{-H}\mathcal{B}'_{+V} \\[.05in]
\mathcal{A}'_{+H}\mathcal{A}'_{-V}\mathcal{B}'_{+H}\mathcal{B}'_{+V},\, \mathcal{A}'_{-H}\mathcal{A}'_{+V}\mathcal{B}'_{+H}\mathcal{B}'_{+V} \\[.05in]
\mathcal{A}'_{+H}\mathcal{A}'_{-V}\mathcal{B}'_{-H}\mathcal{B}'_{-V},\, \mathcal{A}'_{-H}\mathcal{A}'_{+V}\mathcal{B}'_{-H}\mathcal{B}'_{-V} \\[.05in]
\mathcal{A}'_{-H}\mathcal{A}'_{-V}\mathcal{B}'_{+H}\mathcal{B}_{-V}, \, \mathcal{A}'_{-H}\mathcal{A}'_{-V}\mathcal{B}'_{-H}\mathcal{B}'_{+V}. 
\end{array}
\nonumber \\[.01in]
\nonumber 
\end{align}
Here primitive~3's cross-term sign parity is odd, implying that it flips the sign of Eq.~(\ref{densmatstructure})'s off-diagonal elements while leaving its diagonal elements unaffected.  So, with a $\psi^-$ source herald the positive off-diagonal element for $|\gamma|>0$ leads to the $\psi^+$ element in the density matrix's Bell-basis form exceeding its $\psi^-$ element.  We correct this discrepancy by applying a $\pi$ pulse to one of the memories when there is a $\psi^-$ source herald and any of the odd-parity memories heralds.  

At this point we have that Eq.~(\ref{densmatstructure}), with elements given by Eqs.~(\ref{diag11}), (\ref{diag22}), and (\ref{offdiag}), applies to both $\psi^-$ source heralds and all 16 memories heralds.  Focusing now on the $\psi^+$ source heralds we can parallel what we have just done and show that the even-parity memories heralds lead to the same density matrix---before the application of a $\pi$ pulse to one of the memories---as the $\psi^-$ source heralds with the odd-parity memories heralds.  Conversely, the $\psi^+$ source heralds and the odd-parity memories heralds lead to the same density matrix as the $\psi^-$ herald and the even-parity memories heralds.  In conclusion, with judicious use of a $\pi$ pulse,  as shown in Table~\ref{pi_pulse}, all 64 source-memories heralds lead to the loaded-memories' unnormalized density matrix having the structure shown in Eq.~(\ref{densmatstructure}), with elements given by Eqs.~(\ref{diag11}), (\ref{diag22}), and (\ref{offdiag}).

\begin{table}[hbt]
\begin{tabular}{|c|c|c|}\hline
Source Herald & Memories-Herald Parity & Apply $\pi$ pulse  \\[.05in]\hline
$\psi^-$ & even & no \\[.05in]
$\psi^-$ & odd & yes \\[.05in]
$\psi^+$ & even & yes \\[.05in]
$\psi^+$ & odd & no \\ \hline
\end{tabular}
\caption{When to apply a $\pi$ pulse to one of ZALM's quantum memories so that all 64 source-memories heralds lead to the loaded-memories' unnormalized density matrix having the structure shown in Eq.~(\ref{densmatstructure}), with elements given by Eqs.~(\ref{diag11}), (\ref{diag22}), and (\ref{offdiag}).}
\label{pi_pulse}
\end{table} 

\section{Loaded-State Fidelity and Entanglement-Loading Rate \label{FidelityRate}}
\begin{figure}[hbt]
    \centering
\includegraphics[width=0.95\columnwidth]{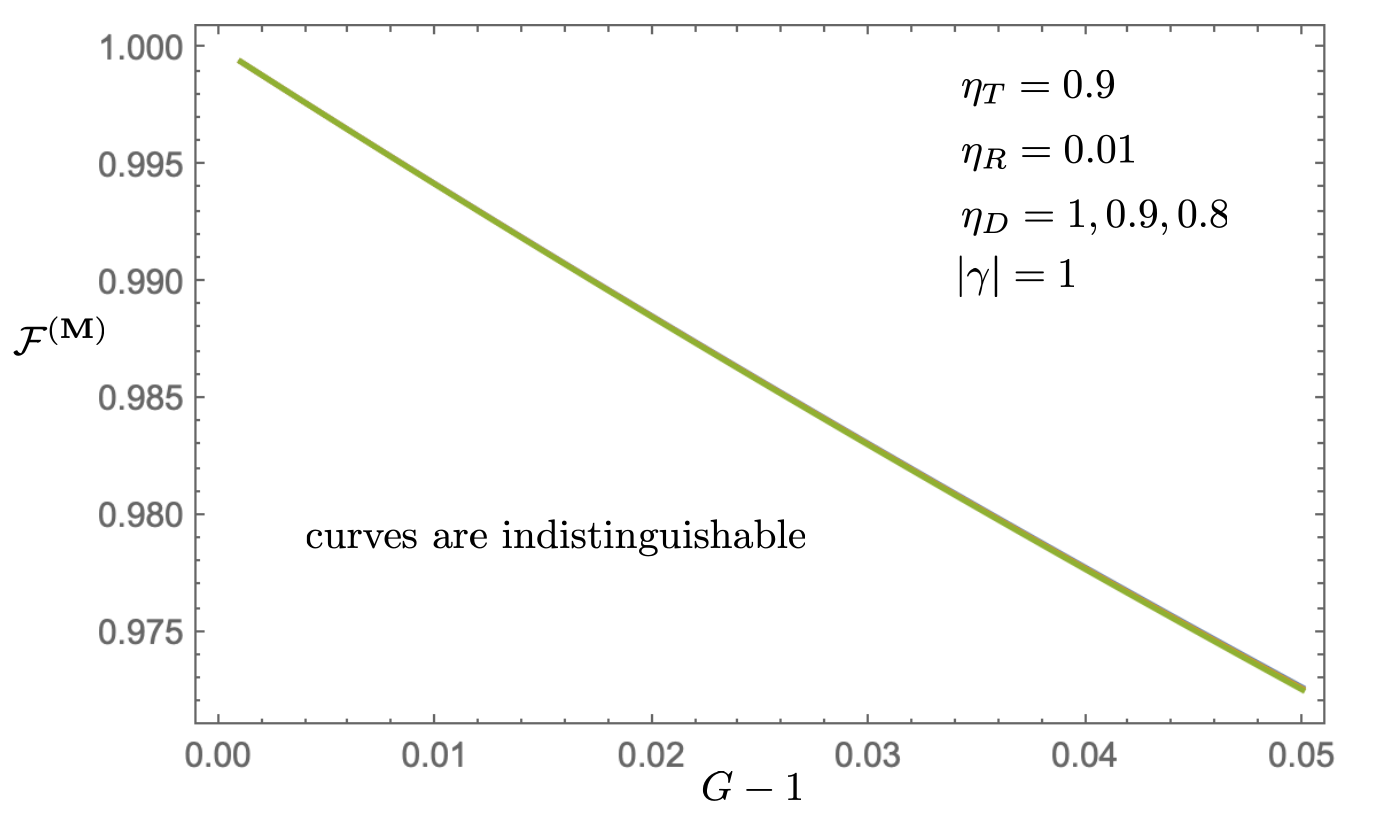}
    \caption{Plots of loaded-state fidelity versus $G-1$, the average number of signal-idler photon pairs per SPDC per pump pulse for the ZALM QTX's two Sagnac sources.  The assumed parameter values are:  90\% QTX partial-BSM efficiency ($\eta_T = 0.9$); 20\,dB QTX-to-QRX propagation loss ($\eta_R = 0.01$); perfect mode matching ($|\gamma|^2 = 1$); and three QRX partial-BSM efficiencies ($\eta_D = 1, 0.9,$ and 0.8).}
 \label{fidelity90} 
\end{figure}

\begin{figure}[hbt]
    \centering
\includegraphics[width=0.95\columnwidth]{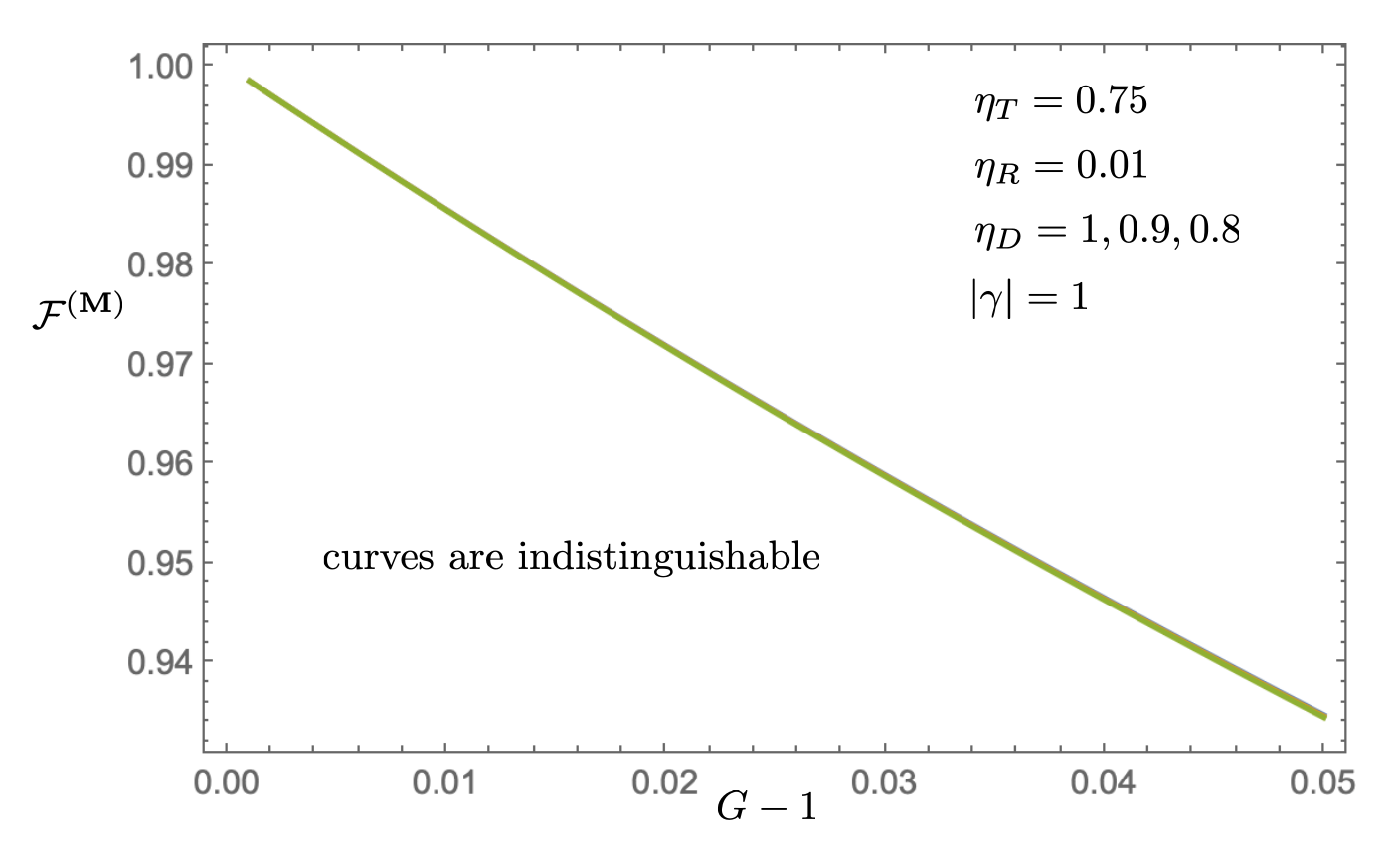}
     \caption{Plots of loaded-state fidelity versus $G-1$, the average number of signal-idler photon pairs per SPDC per pump pulse for the ZALM QTX's two Sagnac sources.  The assumed parameter values are:  75\% QTX partial-BSM efficiency ($\eta_T = 0.75$); 20\,dB QTX-to-QRX propagation loss ($\eta_R = 0.01$); perfect mode matching ($|\gamma|^2 = 1$); and three QRX partial-BSM efficiencies ($\eta_D = 1, 0.9,$ and 0.8).}
 \label{fidelity75} 
\end{figure}

Equations~(\ref{diag11}), (\ref{diag22}), (\ref{offdiag}), and (\ref{rhoB}) provide all the information we need to evaluate fidelity of the loaded-memories' state via,
\begin{equation}
\mathcal{F}^{({\bf M})} \equiv \frac{{}_{\bf M}\langle \psi^-|\tilde{\rho}^{(B)}_{\bf M}|\psi^-\rangle_{\bf M}}{{\rm Tr}[\tilde{\rho}^{(B)}_{\bf M}]} = \frac{\tilde{\rho}_{22}-\tilde{\rho}_{23}}{2(\tilde{\rho}_{11} + \tilde{\rho}_{22})}.
\end{equation}
Figures~\ref{fidelity90} and \ref{fidelity75} plot the fidelity versus $G-1$, the average number of photon pairs generated per SPDC per pump pulse in the ZALM QTX's two Sagnac sources, for 90\% and 75\% QTX partial-BSM efficiencies, respectively.  They both assume perfect mode matching ($|\gamma|^2 = 1$), 20\,dB QTX-to-QRX propagation loss ($\eta_R = 0.01$), and three QRX partial-BSM efficiencies ($\eta_D = 1, 0.9$, and 0.8).  The three curves in each figure are indistinguishable, although not exactly equal because $\mathcal{F}^{({\bf M})}$ \emph{does} depend on $\eta_D$, but only weakly so for the parameter ranges considered in Figs.~\ref{fidelity90} and \ref{fidelity75}.  It is also worth noting that emissive loading is ideal when $|\gamma|^2 =1$ and $\eta_D = 1$, in which case we have
$\mathcal{F}^{({\bf M})} = \mathcal{F}^{({\tilde{\bf S})}}$, i.e., the fidelity of the loaded memories' state equals that of the arriving light~\cite{footnote3}.  

\begin{figure}[hbt]
    \centering
\includegraphics[width=0.975\columnwidth]{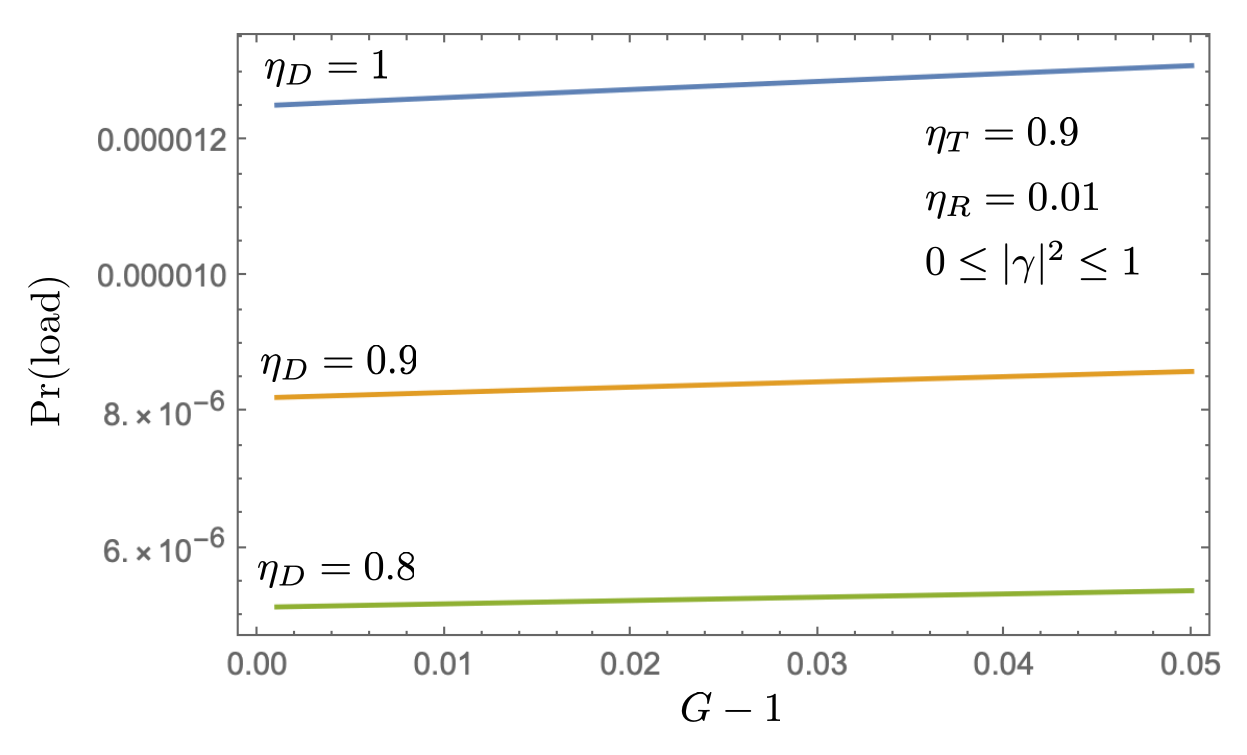}
    \caption{Plots of $\Pr({\rm load})$, the conditional probability of a memories load given there has been a source-memories herald, versus $G-1$, the average number of signal-idler photon pairs per SPDC per pump pulse for the ZALM QTX's two Sagnac sources.  The assumed parameter values are:  90\% QTX partial-BSM efficiency ($\eta_T = 0.9$); 20\,dB QTX-to-QRX propagation loss ($\eta_R = 0.01$); arbitrary mode matching ($0 \le |\gamma|^2 \le 1$); and three QRX partial-BSM efficiencies ($\eta_D = 1, 0.9,$ and 0.8).}
 \label{load90} 
\end{figure}

\begin{figure}[hbt]
    \centering
\includegraphics[width=0.99\columnwidth]{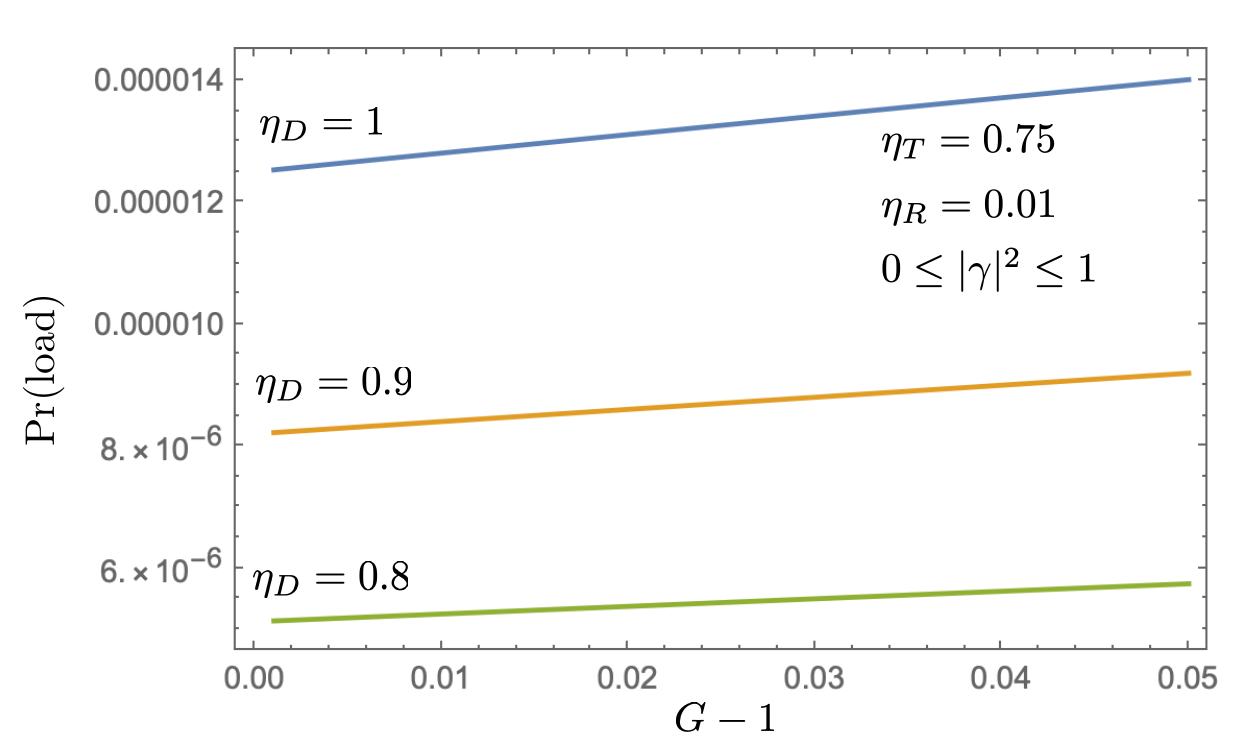}
     \caption{Plots of $\Pr({\rm load})$, the conditional probability of a memories load given there there has been a source-memories herald, versus $G-1$, the average number of signal-idler photon pairs per SPDC per pump pulse for the ZALM QTX's two Sagnac sources.  The assumed parameter values are:  75\% QTX partial-BSM efficiency ($\eta_T = 0.75$); 20\,dB QTX-to-QRX propagation loss ($\eta_R = 0.01$); arbitrary mode matching ($0 \le |\gamma|^2 \le 1$); and three QRX partial-BSM efficiencies ($\eta_D = 1, 0.9,$ and 0.8).}
 \label{load75} 
\end{figure}

Although the loaded-memories' fidelity is almost independent of Alice and Bob's partial-BSM efficiency, $\eta_D$, for the parameters used in Figs.~\ref{fidelity90} and \ref{fidelity75}, such is not the case for $\Pr({\rm load})$, the conditional probability of a memories load given there has been a source-memories herald.  Figures~\ref{load90} and \ref{load75} plot that probability using the $\eta_T, \eta_R$, and $\eta_D$ values assumed in Figs.~\ref{fidelity90} and \ref{fidelity75}, respectively.  Note that these figures apply for \emph{all} $|\gamma|^2$, because $\Pr({\rm load})$ equals the trace of $\tilde{\rho}_{\bf M}$, and the diagonal elements of that matrix are independent of mode mismatch.  More importantly, Figs~\ref{load90} and \ref{load75} show that $\eta_D = 0.8$ reduces the load probability from its $\eta_D = 1$ value by more than 50\%.  

Figures~\ref{load90} and \ref{load75} offer an additional interesting contrast with Figs.~\ref{fidelity90} and \ref{fidelity75} with respect to their dependence on $G-1$.  Fidelity degrades with increasing $G-1$ because, with the assumed 20\,dB propagation loss, a memories load is more likely due to a QTX multipair event, whose load is less likely to be the desired $|\psi^-\rangle_{\bf M}$ state.  On the other hand, with the assumed 20\,dB propagation loss, increasing $G-1$'s increasing the number of QTX multipair events makes it more likely that a loadable event arrives at Alice and Bob's partial-BSM setups.

\begin{figure}[hbt]
    \centering
\includegraphics[width=0.99\columnwidth]{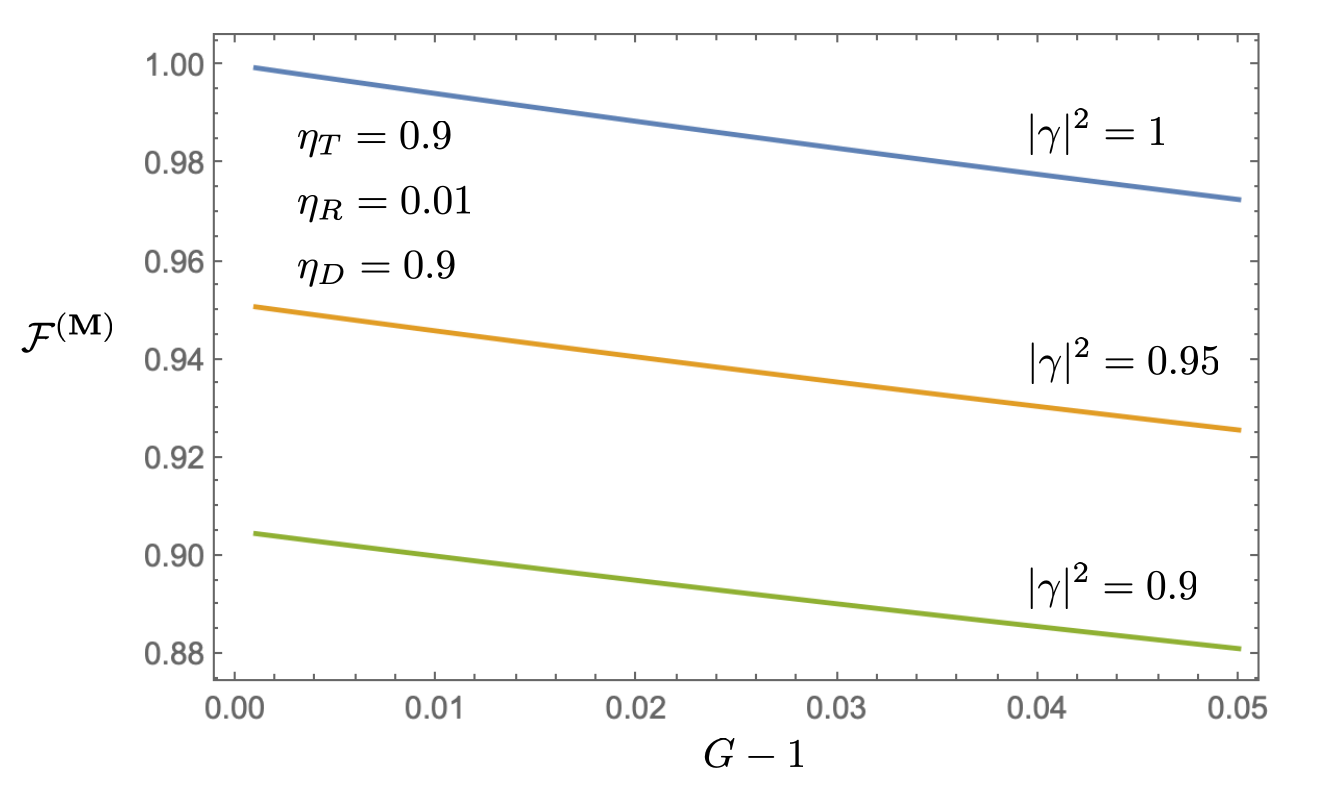}
    \caption{Plots of loaded-state fidelity versus $G-1$, the average number of signal-idler photon pairs per SPDC per pump pulse for the ZALM QTX's two Sagnac sources.  The assumed parameter values are:  90\% QTX partial-BSM efficiency ($\eta_T = 0.9$); 20\,dB QTX-to-QRX propagation loss ($\eta_R = 0.01$); 90\% QRX partial-BSM efficiency ($\eta_D = 0.9$); and three values of mode mismatch ($|\gamma|^2 = 1, 0.95$, and 0.9).} \label{fidmismatch90} 
\end{figure}

\begin{figure}
    \centering
\includegraphics[width=0.975\columnwidth]{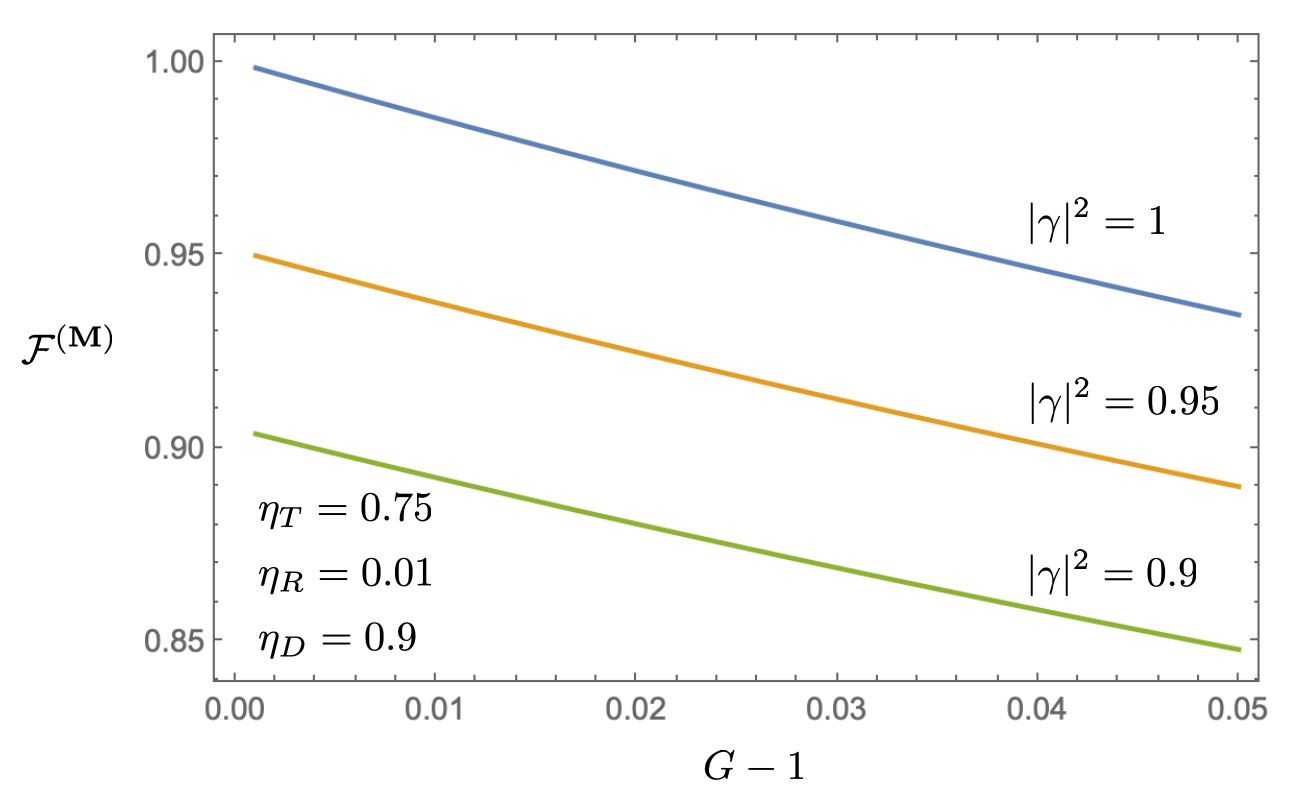}
     \caption{Plots of loaded-state fidelity versus $G-1$, the average number of signal-idler photon pairs per SPDC per pump pulse for the ZALM QTX's two Sagnac sources.  The assumed parameter values are:  75\% QTX partial-BSM efficiency ($\eta_T = 0.75$); 20\,dB QTX-to-QRX propagation loss ($\eta_R = 0.01$); 90\% QRX partial-BSM efficiency ($\eta_D = 0.90$); and three values of mode mismatch ($|\gamma|^2 = 1, 0.95$, and 0.9).}
 \label{fidmismatch75} 
\end{figure}

Figures~\ref{fidmismatch90} and \ref{fidmismatch75} demonstrate the adverse impact of mode mismatch on the loaded-memories' fidelity.  They plot $\mathcal{F}^{({\bf M})}$ versus $G-1$ for $\eta_T = 0.9$ and $\eta_T = 0.75$, respectively, with $\eta_R = 0.01$, $\eta_D = 0.9$ and three values of mode mismatch ($|\gamma|^2 =1, 0.95$, and 0.9).  Roughly speaking, the loaded-memories' fidelity pretty much excludes achieving $\mathcal{F}^{({\bf M})} = 0.99$ unless near-perfect mode matching can be achieved.  

Our final task is assessing the entanglement-loading rates that can be achieved by islands-based ZALM using emissive loading of color-center quantum memories, and comparing them to the rates at which entanglement is delivered to Alice and Bob's QRXs.    To do so we need only combine the loaded-memories' Bell-basis density matrix with the entanglement delivery analysis from Shapiro~\emph{et al}.~\cite{Shapiro2026}.  As a preface to that assessment and comparison, we will summarize some relevant material from Ref.~\cite{Shapiro2026}.

Our entanglement delivery and loading rates for islands-based ZALM make the following assumptions beyond those already made in earlier sections of this paper:  the ZALM QTX's SPDCs have pump-pulse repetition rate $\mathcal{R}_P$; Alice and Bob's QRXs have $N_M \le N_I$ quantum memories; and the ZALM QTX can transmit up to $\mathcal{H}(N_M) \le N_M$ heralds per pump pulse.  From Ref.~\cite{Shapiro2026} we have that
\begin{equation}
\mathcal{R}_E^{({\tilde{\bf S})}} = \mathcal{R}_P\,\mathbb{E}[\mathcal{H}(N_M)]\Pr(|\psi^\pm\rangle_{\tilde{\bf S}}\mid \psi^\pm)
\end{equation}
is the entanglement-delivery rate (in ebits/s) to Alice and Bob's partial-BSM setups. 
Here
\begin{equation}
\mathbb{E}[\mathcal{H}(N_M)] = \sum_{k=0}^{N_M-1}k\,p_k + N_M\sum_{k=N_M}^{N_I}p_k,
\mbox{ for $N_M \le N_I$}, 
\end{equation}
is the average number of SPCI heralds per pump pulse, and $\Pr(|\psi^\pm\rangle_{\tilde{\bf S}}\mid \psi^\pm)$ is the delivery success probability given there has been a source herald.  In the $
\mathbb{E}[\mathcal{H}(N_M)]$ expression, $p_k$ is the probability of having $k$ SPCI heralds and given by
\begin{align}
p_k &= \frac{N_I!}{k!(N_I-k)!}\,q^k(1-q)^{N_I-k} \nonumber \\[.05in]
& \times \left\{\frac{N_I!}{k!(N_I-k)!}\,q^k(1-q)^{N_I-k}\right. 
 \nonumber \\[.05in]  
 & \,\, +2\left.\left[\sum_{\ell = k+1}^{N_I}\frac{N_I!}{\ell!(N_I-\ell)!}\,q^\ell(1-q)^{N_I-\ell}\right]\right\}, 
 \end{align}
 for $k = 0,1,\ldots, N_I$ and 
 \begin{equation}
 q = \frac{2\eta_T(G-1)}{[\eta_T(G-1)+1]^3}.  
 \end{equation}
 Also, the arriving light's success probability is given by
 \begin{widetext}
 \begin{equation}
 \Pr(|\psi^\pm\rangle_{\tilde{\bf S}} \mid \psi^\pm) = 
 \frac{N_S'^4}{2}\!\left[2(1-N_S')^2 -\frac{2\eta_R(3N_S'^3-5N_S'^2+2N_S')}{N_S} + \frac{\eta_R^2(5N_S'^4-6N_S'^3+2N_S'^2)}{N_S^2}\right].
 \end{equation}
 \end{widetext}
 
\begin{figure}[b]
    \centering
\includegraphics[width=0.975\columnwidth]{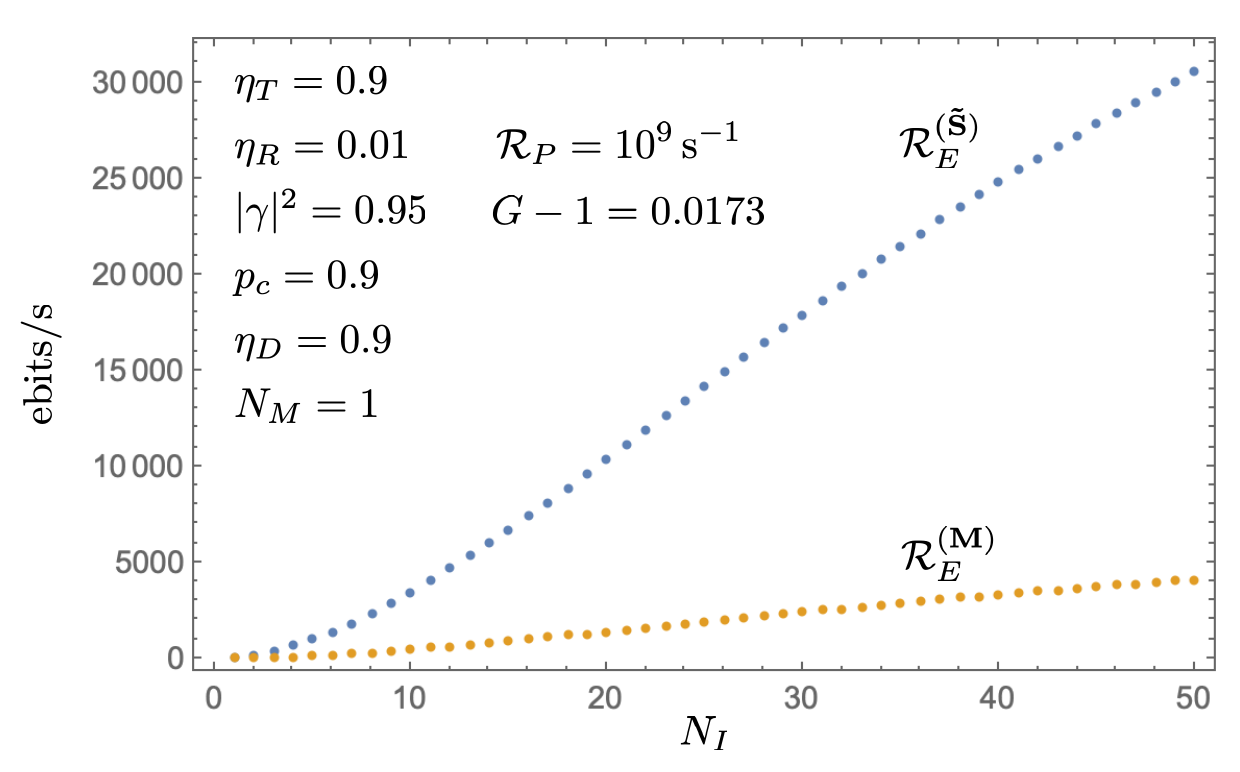}
    \caption{Plots of the single-memory ($N_M = 1$) entanglement delivery and loading rates $\mathcal{R}_E^{({\tilde{\bf S})}}$ and $\mathcal{R}_E^{({\bf M})}$, versus the ZALM QTX's number, $N_I$, of phase-matched spectral islands.  The assumed parameter 
values are:  $\mathcal{R}_P = 10^9\,{\bf s}^{-1}$ pump-pulse repetition rate; $G-1 = 0.0173$ average number of signal-idler photon pairs per SPDC per pump pulse for the ZALM QTX's two Sagnac sources; 90\% QTX partial-BSM efficiency ($\eta_T = 0.9$);  20\,dB QTX-to-QRX propagation loss ($\eta_R = 0.01$); 95\% mode matching ($|\gamma|^2 = 0.95$); 90\% success probability ($p_c = 0.9$) for the memories' emitted photons; and 90\% QRX partial-BSM efficiency ($\eta_D = 0.9$).} \label{ratesNm1} 
\end{figure}

\begin{figure}
    \centering
\includegraphics[width=0.975\columnwidth]{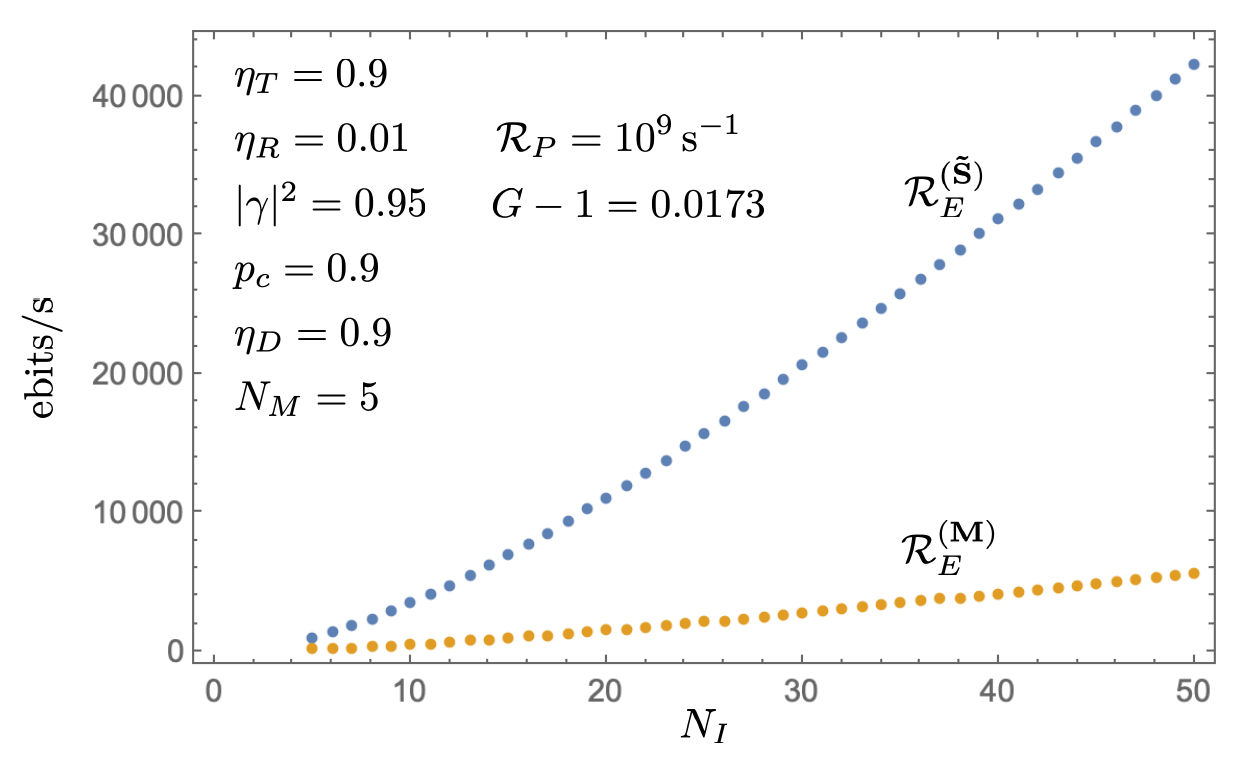}
     \caption{Plots of the multi-memory ($N_M = 5$) entanglement delivery and loading rates $\mathcal{R}_E^{({\tilde{\bf S})}}$ and $\mathcal{R}_E^{({\bf M})}$, versus ZALM's number, $N_I$, of phase-matched spectral islands.  The assumed parameter values are:  $\mathcal{R}_P = 10^9\,{\bf s}^{-1}$ pump-pulse repetition rate; $G-1 = 0.0173$ average number of signal-idler photon pairs per SPDC per pump pulse for the ZALM QTX's two Sagnac sources; 90\% QTX partial-BSM efficiency ($\eta_T = 0.9$);  20\,dB QTX-to-QRX propagation loss ($\eta_R = 0.01$); 95\% mode matching ($|\gamma|^2 = 0.95$); 90\% success probability ($p_c = 0.9$) for the memories' emitted photons; and 90\% QRX partial-BSM efficiency ($\eta_D = 0.9$).} 
 \label{ratesNm5} 
\end{figure}

 The entanglement loading rate---the  ebits/s of successful ($|\psi^-\rangle_{\bf M}$) loads into Alice and Bob's quantum memories---is
 \begin{equation}
 \mathcal{R}_E^{({\bf M)}} = \mathcal{R}_P\,\mathbb{E}[\mathcal{H}(N_M)]p_c^2\Pr(|\psi^-\rangle_{\bf M}),
 \end{equation}
 where $\Pr(|\psi^-\rangle_{\bf M}) = \tilde{\rho}_{22}-\tilde{\rho}_{23}$ is the probability of a $|\psi^-\rangle_{\bf M}$ memory load given there has been a source-memories herald.  
 Figures~\ref{ratesNm1} and \ref{ratesNm5} compare ZALM's entanglement-delivery rate to Alice and Bob's partial-BSM setups with their emissive memories' entanglement-loading rate for single-memory ($N_M =1$) and multi-memory ($N_M = 5$) operation, respectively, as the ZALM QTX's number of phase-matched islands increases.  The $G-1 = 0.0173$ value maximizes the delivery rate for which $\mathcal{F}^{({\tilde{\bf S}})} = 0.99$ when there is 20\,dB QTX-to-QRX propagation loss, while the remaining parameter values are optimistic.  These figures show the beneficial effects of having more phase-matched spectral islands and more quantum memories.  There is, however, a substantial disparity between the entanglement-delivery rate and the entanglement-loading rate.  
 
Emissive loading employs partial BSMs at both Alice and Bob's QRXs, and each partial BSM has 50\% probability of providing a heralding click pattern.  At best, therefore, ZALM with emissive loading will have $\mathcal{R}_E^{(\tilde{\bf M})} = \mathcal{R}_E^{(\tilde{\bf S})}/4$, something that only prevails for ideal ($p_c = 1$, $|\gamma|^2 = 1$, and $\eta_D = 1$) memories loading.  A second difference between the delivered and loaded entanglement is that $|\gamma|^2 = 0.95$ leads to $\mathcal{F}^{({\bf M})} = 0.94$.

\section{Concluding Discussion \label{Discussion}}
At this point we have accomplished the goal laid out in Sec.~\ref{Intro}, namely an end-to-end fidelity and rate analysis for islands-based ZALM that includes:  SPDC multipair emissions of all orders and partial-BSM losses in the ZALM QTX; loss in propagation from the ZALM QTX to Alice and Bob's quantum receivers (QRXs); loss and mode mismatch in the mode converters employed at Alice and Bob's QRXs; and loss in the partial-BSM loading of Alice and Bob's emissive quantum memories.  In this concluding section we will highlight the key takeaway from that analysis, comment on recent related work on islands-based entanglement distribution, use our results to evaluate the hashing bound~\cite{Devetak2005} on our setup's distillable entanglement, and suggest avenues for future research. 

This paper's key takeaway is simple:  islands-based ZALM with emissive-loaded quantum memories suffers a very substantial loss in going from its entanglement-delivery rate to Alice and Bob's partial-BSM setups to the entanglement-loading rate into their memories.  Specifically, Figs.~\ref{ratesNm1} and \ref{ratesNm5} show that islands-based ZALM's entanglement-loading rate is $\sim$0.14 of its entanglement-delivery rate for 20\,dB propagation loss with QRX efficiencies that will likely be quite challenging to achieve.  Much of that rate reduction is due to emissive loading's reliance on partial BSMs, whose efficiencies are 50\% under ideal conditions.  The alternative to emissive loading is Duan-Kimble loading~\cite{Duan2004}, which relies on the state-dependent reflectivities of an intracavity color-center quantum memory~\cite{Raymer2024}.  Unlike emissive loading, which is probabilistic with a 50\% maximum (ideal operation) success probability, Duan-Kimble loading is deterministic.  As discussed below, however, Duan-Kimble loading is not a panacea, but certainly deserving of future study.  Before doing so, we have other topics to discuss.    

Islands-based ZALM and variants thereof have received increasing attention of late~\cite{Chahine2026,Ghormish2026}, and we need to place these works in context with respect to our work in this paper.  Like this paper, both these references assume SPDCs with ideal phase-matched spectral islands.  Both references introduce several alternative heralding architectures for islands-based entanglement distribution and compare their performance to ZALM's with favorable results.  Note that Shapiro~\emph{et al}.~\cite{ShapiroQuantPh} compared ZALM's entanglement-delivery rate to that of  Chahine~\emph{et al}.'s single-Sagnac precursor~\cite{Chahine2025} to their Ref.~\cite{Chahine2026} architectures and found ZALM to be superior.   Unlike our work, whose novelty lies in addressing the entanglement-loading rate to emissive quantum memories, Ref.~\cite{Chahine2026} leaves consideration of memory loading for its  architectures to future work, and, although mentioning the need for mode conversion at Alice and Bob's QRXs, it does not note possible problems arising from mode mismatch.  

Ghormish and Camacho~\cite{Ghormish2026} take a different tack in their approach to islands-based entanglement distribution.  Recognizing that the ZALM QTX's partial BSM has a 50\% chance of producing a false herald caused by one of its Sagnac sources emitting two signal-idler pairs from a single island while the other source's corresponding island emits none, they propose architectures that circumvent this problem by using complementary linear-optics heralding setups that employ a partial BSM and a Green machine.  Doing so, which requires their QTX to employ three or four Sagnac sources instead of ZALM's two, eliminates sending any false heralds to Alice and Bob.  Ghormish and Camacho do present results for the entanglement-loading rate into quantum memories at Alice and Bob's QRXs, but they do not specify whether the memories were emissively or Duan-Kimble loaded.  Also, like Chahine~\emph{et al}.~\cite{Chahine2026}, Ghormish and Camacho do not note the possibility of mode mismatch between the arriving light and the quantum memories.  Consequently, to the best of our knowledge, our end-to-end fidelity and rate analysis is the first to include mode mismatch and emissive-memory loading in its treatment of islands-based entanglement distribution.  

\begin{figure}[h]
    \centering
\includegraphics[width=0.975\columnwidth]{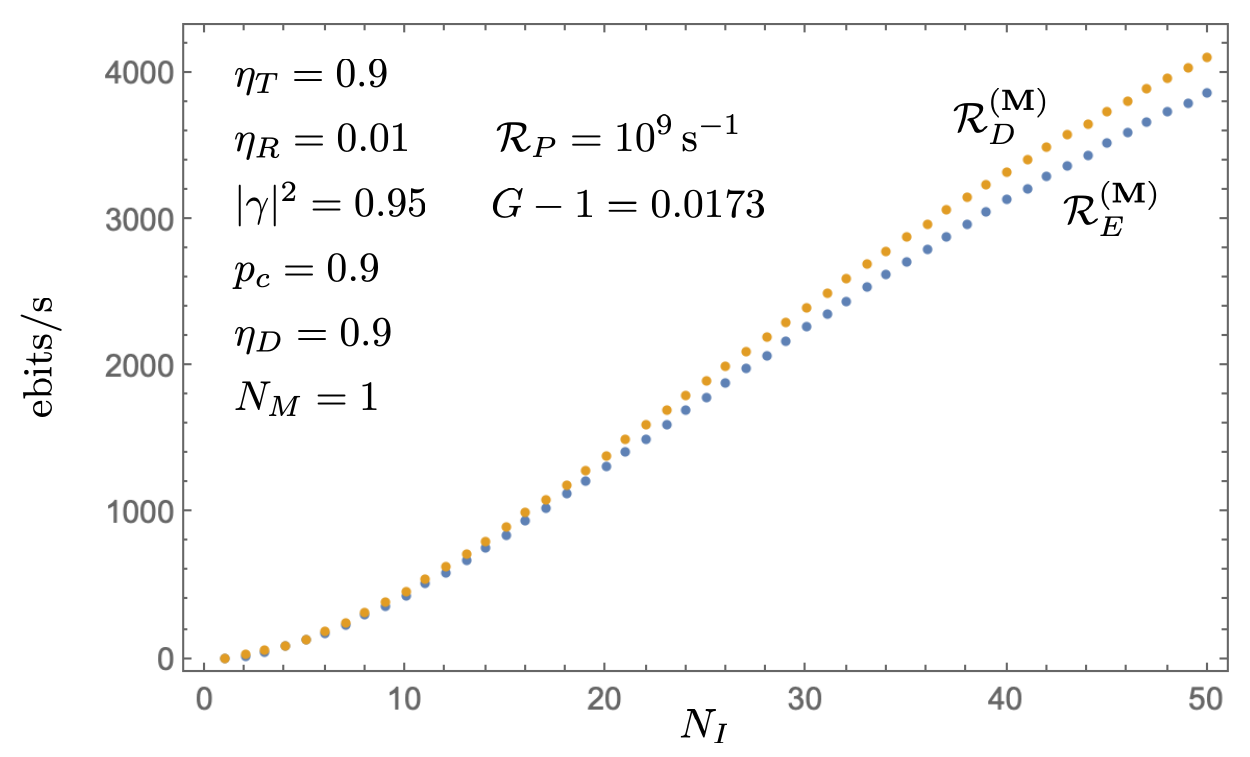}
    \caption{Plots of the single-memory ($N_M = 1$) entanglement-loading rate, $\mathcal{R}_E^{({\bf M})}$, and the hashing bound on the  distillable-entanglement rate, $\mathcal{R}_D^{({\bf M})}$, versus the ZALM QTX's number, $N_I$, of phase-matched spectral islands.  The assumed parameter values are:  $\mathcal{R}_P = 10^9\,{\rm s}^{-1}$, $G-1 = 0.0173$, $\eta_T = 0.9$, $\eta_R = 0.01$, $|\gamma|^2 = 0.95$, $p_c = 0.9$, and $\eta_D = 0.9$.} \label{HashNm1} 
\end{figure}

\begin{figure}[h]
    \centering
\includegraphics[width=0.975\columnwidth]{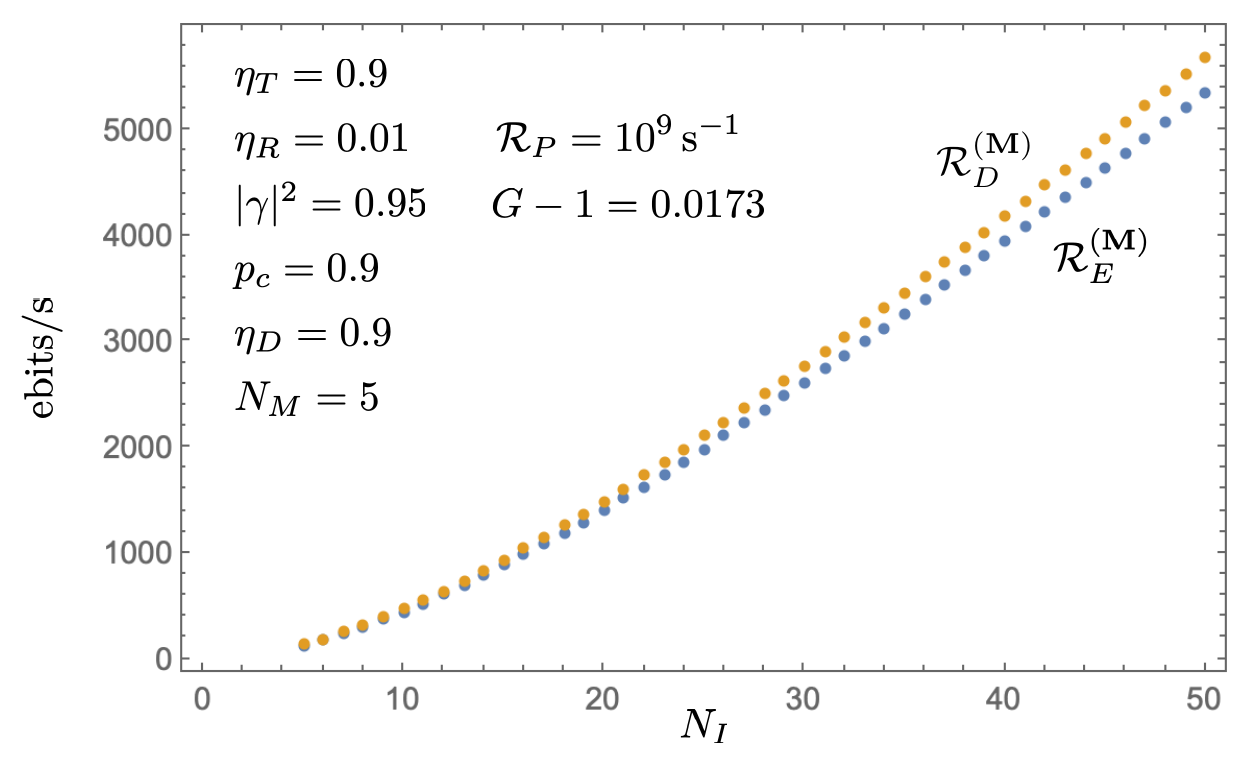}
     \caption{Plots of the multi-memory ($N_M = 5$) entanglement-loading rate, $\mathcal{R}_E^{({\bf M})}$, and the hashing bound on the  distillable-entanglement rate, $\mathcal{R}_D^{({\bf M})}$, versus the ZALM QTX's number, $N_I$, of phase-matched spectral islands.  The assumed parameter values are:  $\mathcal{R}_P = 10^9\,{\rm s}^{-1}$, $G-1 = 0.0173$, $\eta_T = 0.9$, $\eta_R = 0.01$, $|\gamma|^2 = 0.95$, $p_c = 0.9$, and $\eta_D = 0.9$.} 
 \label{HashNm5} 
\end{figure}

Thus far we have not addressed the transmission rate of distillable entanglement loaded into Alice and Bob's quantum memories.  In particular, Sec.~\ref{FidelityRate}'s $\mathcal{R}_E^{({\bf M})}$ is the number of $|\psi^-\rangle_{\bf M}$ states/sec loaded into Alice and Bob's quantum memories, but, for parameter values of interest, the loaded states are \emph{mixed} states.  Devetak and Winter's hashing bound~\cite{Devetak2005} provides a convenient achievable lower bound on the distillable-entanglement rate, which is given by
\begin{align}
\mathcal{R}_D^{({\bf M})}  &= \mathcal{R}_P\,\mathbb{E}[\mathcal{H}(N_M)]\Pr({\rm load})
\nonumber \\[.05in]
&\,\,\times
\{S(\hat{\rho}_{\bf M}) -\max[S(\hat{\rho}_{M_A}),S(\hat{\rho}_{M_B})]\},
\end{align}
where $S(\hat{\rho})$ is the von Neumann entropy of $\hat{\rho}$.  Because $\hat{\rho}_{\bf M}$ is diagonal in the Bell basis we get
\begin{align}
\mathcal{R}_D^{({\bf M})} &= \mathcal{R}_P\,\mathbb{E}[\mathcal{H}(N_M)]\Pr({\rm load}) \nonumber \\[.05in]
& \,\,\times \left\{1-\sum_{s=-,+}\Pr(|\psi^s\rangle_{\bf M})\log_2[\Pr(|\psi^s\rangle_{\bf M})]\right. \nonumber \\[.05in]
&\,\,- \left.\sum_{s=+,-}\Pr(|\phi^s\rangle_{\bf M})\log_2[\Pr(|\phi^s\rangle_{\bf M})] \right\}.
\end{align}
Figures~\ref{HashNm1} and \ref{HashNm5} compare ZALM's entanglement-loading rate to its distillable-entanglement rate's hashing bound for single-memory ($N_M =1$) and multi-memory ($N_M = 5$) operation, respectively, as the ZALM QTX's number of phase-matched islands increases.  These figures employ the parameter values used in Figs.~\ref{ratesNm1} and \ref{ratesNm5}:  $\mathcal{R}_P = 10^9\,{\rm s}^{-1}$, $G-1 = 0.0173$, $\eta_T = 0.9$, $\eta_R = 0.01$, $|\gamma|^2 = 0.95$, $p_c = 0.9$, and $\eta_D = 0.9$.  Because 
$S(\hat{\rho}_{\bf M}) -\max[S(\hat{\rho}_{M_A}),S(\hat{\rho}_{M_B})] > 0.9999$ for these parameter values, the distillable-entanglement rate is extremely close to the unachievable (for mixed-state $\hat{\rho}_{\bf M}$'s) upper bound on the distillable-entanglement rate,
\begin{equation}
\mathcal{R}_U^{({\bf M})}  = \mathcal{R}_P\,\mathbb{E}[\mathcal{H}(N_M)]\Pr({\rm load}),
\end{equation}
which assumes that the loaded state is a pure $|\psi^-\rangle_{\bf M}$ state.  This confluence is directly attributable to the assumed parameter values giving a high-fidelity, $\mathcal{F}^{({\bf M})} = 0.94$, loaded state.  

We will wrap up our discussion with some suggestions for future work.  Shapiro~\emph{et al}.~\cite{Shapiro2026} compared the single-memory and multi-memory entanglement-delivery rates for islands-based ZALM with their unheralded single-SPDC counterparts with the goal of answering the ``to herald or not to herald'' question.  A natural first follow-on to our present work is therefore to extend that comparison to the entanglement-loading rates for systems using emissive memories.  

Earlier we noted that a substantial rate loss between ZALM's entanglement-delivery performance and its entanglement-loading performance was incurred with emissive memories owing their combination of probabilistic operation and non-idealities.  There we suggested the deterministic, Duan-Kimble, loading might prove advantageous in this regard.     In previous ZALM work~\cite{Shapiro2024,Raymer2024} we presumed DWDM-channelization of conventional ZALM Sagnac sources operating in the perturbative regime wherein multipair emissions are negligible. There we showed that, except for some rate loss, perfect state transfer from the mode-converted arriving light to the quantum memories was achievable when the intracavity quantum memories' state-dependent reflectivities were constant over the bandwidth of the mode-converted arriving light.  Moreover, the perfect state transfer in this narrowband regime did \emph{not} require the mode-converted light to be in a single temporal mode, obviating emissive loading's onerous mode-matching burden.  Duan-Kimble loading has its own onerous mode-conversion burden:  compressing the wavelength-converted arriving light's bandwidth to well within the $\sim$100\,MHz bandwidth of an intracavity quantum memory.  

It would be nice to lift Refs.~\cite{Shapiro2024,Raymer2024}'s restriction to the perturbative regime and complete a Duan-Kimble loading analysis for islands-based ZALM that accounts for multipairs of all orders.  Unfortunately, that would require evaluating the state-dependent reflectivities for multipair arriving light, and, as yet, no analytic results are available for those multipair reflectivities.  Nevertheless, because the arriving light's Bell-state fraction, for the parameter values assumed in Figs.~\ref{HashNm1} and \ref{HashNm5}, exceeds 0.999~\cite{Shapiro2026}, a reasonable approach to obtaining Duan-Kimble loading results for islands-based ZALM is to discard the multipair contribution to the arriving-light's density operator and proceed with the transform-domain analysis along the lines used in this paper for emissive loading.   Performing that analysis is our second suggestion for future work.  A third suggestion would then be to address ``to herald or not herald'' for Duan-Kimble loading.  

Our last suggestion for future work is to apply our emissive-memory loading analysis to the alternative entanglement-distribution architectures proposed in  Refs.~\cite{Chahine2026,Ghormish2026}.  

\acknowledgments
This work was supported by the Engineering Research Centers Program of the National Science Foundation
under Grant \#1941583 to the NSF-ERC Center for Quantum Networks. The author thanks Saikat Guha for encouraging him to incorporate emissive memories into ZALM analysis, and J. Gabriel Richardson for information about such memories.


\begin{thebibliography}{99}
\bibitem{footnote0}Collinear operation ensures the Sagnac sources' signal-idler photon pairs will be in single spatial modes, as needed for efficient QTX-to-QRXs transmission whether those links be through free space or over optical fiber.
\bibitem{footnote1}Strictly speaking, the $|\psi\rangle_{S_kI_k}$ only specify the joint states of their signal and idler's single temporal-mode excitations; all their other temporal modes are in their vacuum states.  Moreover, as discussed below, no loss of generality is entailed by assuming the $|\psi\rangle_{S_kI_k}$ are pure states.
\bibitem{footnote2}Mode shaping will be needed if, for example, $\phi_S(t)$ is a Gaussian mode but $\phi_E(t)$ is a Lorentzian mode.
\bibitem{footnote3}Remember, we have defined the ``arriving light'' to be the mode-converted signal photons arriving at Alice and Bob's partial-BSM setups.  Thus losses in Alice and Bob's mode converters have been lumped into the propagation loss.  Likewise, losses in coupling their memories' emitted photons to those setups are accounted for by their success probabilities, $p_c$, that will figure in our rate analysis but do not impact loaded-state fidelity as that assumes there has been a memories herald.
\end{thebibliography}
\end{document}